\documentstyle[epsfig,12pt]{article}
\newcommand{\la}[1]{\label{#1}}
\newcommand{\ur}[1]{(\ref{#1})}

\newcommand{\eq}[1]{eq.~(\ref{#1})}
\newcommand{\eqs}[2]{eqs.~(\ref{#1}, \ref{#2})}
\newcommand{\Eq}[1]{Eq.~(\ref{#1})}

\newcommand{\e}{\epsilon}

\def\Tr{\mbox{Tr}}
\def\beq{\begin{equation}}
\def\eeq{\end{equation}}
\def\bea{\begin{eqnarray}}
\def\eea{\end{eqnarray}}

\begin{document}
\thispagestyle{empty}
\begin{flushright} NORDITA-99/85 HE \end{flushright}
\vskip 2true cm
\begin{center}
{\Large\bf Yang--Mills Theory in Three Dimensions \\
\vskip .3true cm
as Quantum Gravity Theory} \\

\vskip 1.5true cm

{\large\bf Dmitri Diakonov$^{\diamond *}$ and Victor Petrov}$^*$ \\
\vskip 1true cm
$^\diamond$ {\it NORDITA, Blegdamsvej 17, DK-2100 Copenhagen \O,
Denmark} \\
\vskip .5true cm
$^*$ {\it Petersburg Nuclear Physics Institute, Gatchina,
St.Petersburg 188350, Russia} \\
\vskip .5true cm
E-mail: diakonov@nordita.dk, victorp@thd.pnpi.spb.ru
\end{center}
\vskip 1.5true cm
\begin{abstract}
\noindent
We perform the dual transformation of the Yang--Mills theory in $d=3$
dimensions using the Wilson action on the cubic lattice. The dual
lattice is made of tetrahedra triangulating a 3-dimensional curved
manifold but embedded into a flat 6-dimensional space (for the $SU(2)$
gauge group). In the continuum limit the theory can be reformulated
in terms of 6-component gauge-invariant scalar fields having the meaning
of the external coordinates of the dual lattice sites. These 6-component
fields induce a metric and a curvature of the 3-dimensional dual colour
space. The Yang--Mills theory can be identically rewritten as a quantum
gravity theory with the Einstein--Hilbert action but purely imaginary
Newton constant, plus a homogeneous `matter' term. Interestingly,
the theory can be formulated in a gauge-invariant and local form without
explicit colour degrees of freedom.
\end{abstract}

\newpage

\section {Lattice partition function}

Though our objective is the continuum theory we start by formulating
the $SU(N_c)$ gauge theory on a cubic lattice. The partition function
can be written as an integral over all link variables being $SU(N_c)$
unitary matrices $U$ with the action being a sum over plaquettes,

\beq
{\cal Z}(\beta)=\int \prod_{{\rm links}} dU_{{\rm link}}
\exp\left(\sum_{{\rm plaquettes}}\beta\;(\Tr\;
U_{{\rm plaq}}+{\rm c.c.})\;/\;2\,\Tr\;1\right)
\la{Z1}\eeq
where $\beta$ is the dimensionless inverse coupling. The unitary
matrix $U_{{\rm plaq}}$ is a product of four link unitary matrices
closing a plaquette.

To get to the continuum limit one writes $U_{{\rm link}}=
\exp(iaA_\mu^at^a)$ where $a$ is the lattice spacing and $A_\mu^at^a
=A_\mu$ is the Yang--Mills gauge potential with $t^a$ being the
generators of the gauge group normalized to $\Tr\;t^at^b=\delta^{ab}/2$,
and expands $\Tr\;U_{{\rm plaq}}$ in the lattice spacing $a$. As a
result one gets for a plaquette lying in the $(12)$ plane:

\beq
\beta\frac{\Tr\;U_{{\rm plaq}}+{\rm c.c.}}{2\;\Tr\;1}
=\beta\left(1-a^4\frac{\Tr\;F_{12}^2}{2\,\Tr\;1}+O(a^6)\right),
\la{TrU}\eeq
where $F_{\mu\nu}=\partial_\mu A_\nu-\partial_\nu A_\mu-i
[A_\mu A_\nu]$ is the Yang--Mills field strength. Summing over all
plaquettes one obtains the partition function of the continuum theory,

\beq
{\cal Z}_{{\rm cont}}=\int DA_\mu\exp\left(-\frac{1}{2g_d^2}
\int d^dx\;\Tr\;F_{\mu\nu}^2\right),
\la{Z2}\eeq
with an obvious relation between the dimensionless lattice
coupling $\beta$ and the $SU(N_c)$ gauge coupling constant in $d$
dimensions, $g_d^2$:

\beq
\beta=\frac{2N_c}{a^{4-d}g_d^2}.
\la{betagen}\eeq

In this paper we concentrate on the Euclidean $SU(2)$ Yang--Mills
theory in $d=3$ dimensions. In this case \eq{betagen} reads:

\beq
\beta=\frac{4}{ag_3^2}.
\la{beta}\eeq

The continuum limit of the $d=3$ Yang--Mills theory given by
the partition function \ur{Z1} is obtained as one takes the lattice
spacing $a\to 0$ and $\beta\to\infty$ with their product
$g_3^2=4/(a\beta)$ fixed. This quantity provides the theory with a
mass scale. It is widely believed (though not proven so far) that the
theory possesses two fundamental properties:  1) the average of a large
Wilson loop has an area behaviour with a string tension proportional to
$g_3^4$, 2) correlation functions of local operators like
$F_{\mu\nu}^2$ decay exponentially at large separations, with a `mass
gap' proportional to $g_3^2$.

Our aim will be to rewrite the partition function \ur{Z1} in dual
variables and to study its continuum limit.

\section{Dual transformation}

The general idea is to integrate over link variables $U_{{\rm link}}$
in \eq{Z1} and to make a Fourier transformation in the plaquette
variables $U_{{\rm plaq}}$. This will be made in several steps, one for
a subsection.

\subsection{Inserting a unity into the partition function}

First of all, one needs to introduce explicitly integration over
unitary matrices ascribed to the plaquettes, $U_{{\rm plaq}}$. This is
done by inserting a unity for each plaquette into the partition
function \ur{Z1}:

\beq
1=\prod_{{\rm plaquettes}}\int dU_{{\rm plaq}}\;
\delta(U_{{\rm plaq}}\,,\;U_1U_2U_3U_4)
\la{unity}\eeq
where $U_{1...4}$ are the link variables closing into a given
plaquette. The $\delta$-function is understood with the
group-invariant Haar measure. A realization of such a
$\delta$-function is given by Wigner $D$-functions:

\beq
\delta(U,V)=\sum_{J=0,\frac{1}{2},1,\frac{3}{2},...}
(2J+1)D^J_{m_1m_2}(U^\dagger)D^J_{m_2m_1}(V).
\la{delta}\eeq
This equation is known as a completeness condition for the
$D$-functions \cite{MVKh}. The main properties of the $D$-functions
used in this paper are listed in Appendix A.

\Eq{delta} should be understood as follows: if one integrates any
function of a unitary matrix $U$ with the r.h.s. of \eq{delta} over the
Haar measure $dU$ one gets the same function but of the argument $V$:

\beq
\int dU\;f(U)\;\delta(U,V)=f(V).
\la{deltaf}\eeq

Using the multiplication law for the $D$-functions (see Appendix A,
\eq{multiplication}) one can write down the unity to be inserted for
each plaquette in the partition function \ur{Z1} as

\beq
1=\int dU_{{\rm plaq}}\;\sum_J(2J+1)\;D^J_{m_1m_2}(U_{{\rm
plaq}}^\dagger) D^J_{m_2m_3}(U_1)D^J_{m_3m_4}(U_2)
D^J_{m_4m_5}(U_3) D^J_{m_5m_1}(U_4)
\la{unity2}\eeq
where $U_{1...4}$ are the corresponding link variables forming
the plaquette under consideration.

\subsection{Integrating over plaquette variables}

Integrating over plaquette unitary matrices $U_{{\rm plaq}}$ becomes
now very simple. For each plaquette of the lattice one has
factorized integrals of the type

\beq
\int dU_{{\rm plaq}}\;\exp\left(\beta\frac{\Tr\;U_{{\rm plaq}}+
\Tr\;U_{{\rm plaq}}^\dagger}{2\,\Tr\;1}\right)\;
D^J_{m_1m_2}(U_{{\rm plaq}}^\dagger)
=\delta_{m_1m_2}\frac{2}{\beta}I_1(\beta)T_J(\beta),
\la{plaq1}\eeq
where $T_J(\beta)$ is the ratio of the modified Bessel functions
\cite{DZ},

\beq
T_J(\beta)=\frac{I_{2J+1}(\beta)}{I_1(\beta)}\;\longrightarrow\;
\exp\left[-\frac{2J(J+1)}{\beta}\right]\;{\rm at}\;\beta\to\infty.
\la{T_J}\eeq
The quantity $T_J(\beta)$ is the `Fourier transform' of the Wilson
action; since in the lattice formulation the dynamical variables have
the meaning of Euler angles and are therefore compact, the Fourier
transform depends on discrete values $J=0,\,1/2,\,1,\,3/2,...$.
However, as one approaches the continuum limit ($\beta\to\infty$) the
essential values of the plaquette angular momenta increase as
$J\sim\surd{\beta}$ and their discreteness becomes less relevant.
Strictly speaking, the continuum limit is achieved at plaquette angular
momenta $J\gg 1$.

We would like to make a side remark on this occasion. The quantity
$T_J(\beta)$ gives the probability that plaquette momenta $J$ is
excited, for given $\beta$. For a typical value used in lattice
simulations $\beta=2.6$ (in 4 dimensions) we find that the
probabilities of having plaquette excitations with
$J=0,\,1/2,\,1,\,3/2\;{\rm and}\;2$ are 56\%, 29\%, 11\%, 3\% and 1\%,
respectively. It means that lattice simulations are actually dealing
mainly with $J=0,\,1/2\;{\rm and}\;1$ with a tiny admixture of higher
excitations. It would be important to understand why and how
continuum physics is reproduced by lattice simulations despite only
such small values of plaquette $J$'s are involved.

We get, thus, for the partition function:

\[
{\cal Z}=\left[\frac{2}{\beta}I_1(\beta)\right]^{\#{\rm\;of\;plaquettes}}
\sum_{J_P}\quad\prod_{{\rm plaquettes}}(2J_P+1)\;T_{J_P}(\beta)
\]
\beq
\times\prod_{{\rm links}\;l}\int dU_l\;
D^{J_P}_{m_1m_2}(U_1)\;D^{J_P}_{m_2m_3}(U_2)\;
D^{J_P}_{m_3m_4}(U_3)\;D^{J_P}_{m_4m_1}(U_4)
\la{Z3}\eeq
where $U_{1\!-\!4}$ are link variables forming a plaquette with angular
momentum $J_P$.

\subsection{Integrating over link variables}

The difficulty in performing integration over link variables in \eq{Z3}
is due to the fact that any link enters several plaquettes.
In $d=2$ dimensions every link is shared by two plaquettes, hence one
has to calculate integrals of the type

\beq
\int dU D^{J_1}_{kl}(U) D^{J_2}_{mn}(U^\dagger)
=\frac{1}{2J_1+1}\delta_{J_1J_2}\;\delta_{kn}\;\delta_{lm}
\la{d2link}\eeq
for all links on the lattice. We shall consider this case later,
in section 4.

In $d=3$ dimensions every link is shared by four plaquettes, hence the
integral over link variables is of the type

\beq
\int dU D^{J_A}_{m_1m_2}(U)D^{J_B}_{m_3m_4}(U)
D^{J_C}_{m_5m_6}(U)D^{J_D}_{m_7m_8}(U)
\la{d3link}\eeq
where $J_{A,B,C,D}$ are angular momenta associated with four plaquettes
intersecting at a given link $U$, and $m_{1\!-\!8}$ are `magnetic'
quantum numbers, to be contracted inside closed plaquettes. In $d=4$
dimensions there will be six plaquettes intersecting at a given link
but we shall not consider this case here.

The general strategy in calculating the link integrals \ur{d3link} is
(i) to divide by a certain rule four $D$-functions into two pairs
and to decompose the pairs of $D$-functions in terms of single
$D$-functions using \eq{decomposition} of Appendix A, (ii) to
integrate the resulting two $D$-functions using \eq{d2link} and,
finally, (iii) to contract the `magnetic' indices. Since all
`magnetic' indices will be eventually contracted we shall arrive
to the partition function written in terms of the invariant $3nj$
symbols.

There are several different tactics how to divide four $D$-functions
into two pairs, eventually leading to anything from $6j$ to $18j$
symbols. In this paper we take a route used in refs.\cite{ACS,HS},
leading to a product of many $6j$ symbols, although on this route one
looses certain symmetries, and that causes difficulties later on.
The gain, however, is that it is more easy to work with $6j$ symbols
than with $12j$ or $18j$ symbols. Since important sign factors have been
omitted in refs.\cite{ACS,HS} and only final result has been reported
there, we feel it necessary to give a detailed derivation below.

\begin{figure}
\centerline{
\epsfxsize 6.0cm
\epsfysize 6.5cm
\epsffile{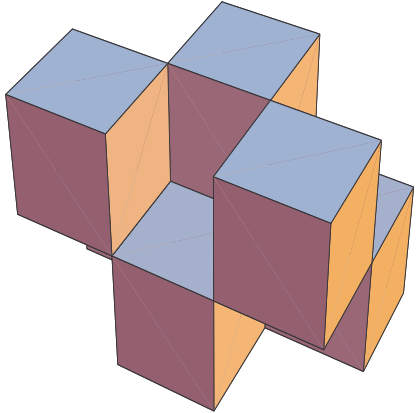}}
\caption[]{`Even' cubes in checker board order.}
\end{figure}

In $d=3$ dimensions all plaquettes are shared by two adjacent cubes,
therefore, it is natural to divide all cubes of the lattice into
two classes which we shall call `even' and `odd', and to
attribute plaquettes only to even cubes. We shall call the cube
even if its left-lower-forward corner is a lattice site with even
coordinates, $(-1)^{x+y+z}=+1$. It will be called odd in the
opposite case.  The even and odd cubes form a 3-dimensional
checker board, as illustrated in Fig.1, where only even cubes are
drawn explicitly.  The even cubes touch each other through a common
edge or link, as do the odd ones among themselves. The even
and odd cubes have common faces or plaquettes. All plaquettes
will be attributed to even cubes only:  that is the reason for the
division of cubes into two classes.

Let us consider an  even cube shown in Fig.2.

\begin{figure}
\begin{picture}(100,150)(0,0)
\epsfxsize 10.0cm
\epsfysize 10.0cm
\put(80,-42){
\epsfbox{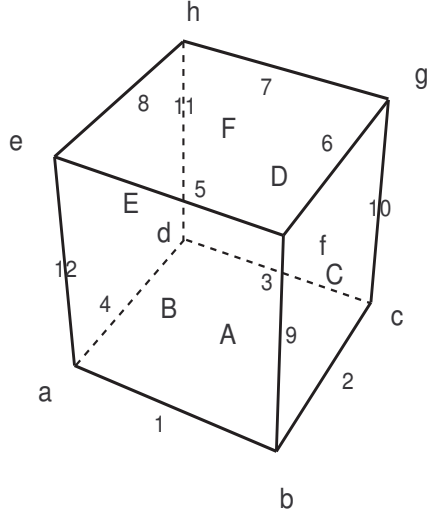}
}
\end{picture}
\caption[]{Elementary even cube}
\end{figure}

$A,B,C,D,E,F$ denote its 6 faces, numbers from 1 to 12 denote its links
or edges, $a,b,c,d,e,f,g,h$ denote its 8 vertices or sites.
Correspondingly, we shall denote plaquette angular momenta by
$J_{A\!-\!F}$, link variables by $U_{1\!-\!12}$, and the `magnetic'
numbers of the $D$-functions will carry indices $a\!-\!h$ referring to
the sites the $D$-functions are connecting.

One can write the traces of products of four $D$-functions over
plaquettes in various ways.  To be systematic we shall adhere to the
following rule: Link variables in the plaquette are taken in the
anti-clock-wise order, {\it as viewed from the center of the
even cube to which the given plaquette belongs.} If the link goes
in the positive direction of the $x,y,z$ axes we ascribe the $U$
variable to it; otherwise we ascribe the $U^\dagger$ variable to it.

With these rules the six plaquettes of the elementary cube shown in
Fig.2 bring in the following six traces of the $D$-function products:

\[{\rm Cube}=
\left[D^{J_A}_{i_ai_b}(U_1) D^{J_A}_{i_bi_c}(U_2)
D^{J_A}_{i_ci_d}(U_3^\dagger)D^{J_A}_{i_di_a}(U_4^\dagger)\right]\;
\left[D^{J_B}_{j_bj_a}(U_1^\dagger) D^{J_B}_{j_aj_e}(U_{12})
D^{J_B}_{j_ej_f}(U_5)D^{J_B}_{j_fj_b}(U_9^\dagger)\right]
\]
\[
\left[D^{J_C}_{k_ck_b}(U_2^\dagger) D^{J_C}_{k_bk_f}(U_9)
D^{J_C}_{k_fk_g}(U_6)D^{J_C}_{k_gk_c}(U_{10}^\dagger)\right]\;
\left[D^{J_D}_{l_dl_c}(U_3) D^{J_D}_{l_cl_g}(U_{10})
D^{J_D}_{l_gl_h}(U_7^\dagger)D^{J_D}_{l_hl_d}(U_{11}^\dagger)\right]
\]
\beq
\left[D^{J_E}_{m_em_a}(U_{12}^\dagger) D^{J_E}_{m_am_d}(U_4)
D^{J_E}_{m_dm_h}(U_{11})D^{J_E}_{m_hm_e}(U_8^\dagger)\right]\,
\left[D^{J_F}_{n_fn_e}(U_5^\dagger) D^{J_F}_{n_en_h}(U_8)
D^{J_F}_{n_hn_g}(U_7)D^{J_F}_{n_gn_f}(U_6^\dagger)\right].
\la{cube1}\eeq

Each link variable $U_{1\!-\!12}$ appears in this product twice:
once as $U$, the other time as $U^\dagger$. For $D(U^\dagger)$ we use
\eq{phase} of the Appendix A to write it in terms of $D(U)$. After that
we can apply the decomposition rule \ur{decomposition} of that Appendix
to write down pairs of $D$-functions in terms of one $D$-function
and two $3jm$ symbols. The new $D$-functions correspond to the {\em
links} and carry angular momenta which we denote by $j$'s. The $3jm$
symbols have `magnetic' indices which get contracted when all indices
related to a given corner of the cube are assembled together. Though
this exercise is straightforward it is rather lengthy, and we relegate
it to Appendix B.  As a result we get the following expression which is
identically equal to \ur{cube1}:

\[{\rm Cube}=
\sum_{j_1...j_{12}}(2j_1+1)...(2j_{12}+1)
\]
\[
D^{j_1}_{o_ao_b}(U_1)
D^{j_2}_{-p_c,-p_b}(U_2^\dagger)
D^{j_3}_{-q_c,-q_d}(U_3^\dagger)
D^{j_4}_{r_ar_d}(U_4)
D^{j_5}_{-s_f,-s_e}(U_5^\dagger)
D^{j_6}_{t_ft_g}(U_6)
\]
\[
D^{j_7}_{u_hu_g}(U_7)
D^{j_8}_{-v_h,-v_e}(U_8^\dagger)
D^{j_9}_{-w_f,-w_b}(U_9^\dagger)
D^{j_{10}}_{x_cx_g}(U_{10})
D^{j_{11}}_{-y_h,-y_d}(U_{11}^\dagger)
D^{j_{12}}_{z_az_e}(U_{12})
\]
\[
\left(\begin{array}{ccc}j_1&j_4&j_{12}\\o_a&r_a&z_a\end{array}\right)
\left(\begin{array}{ccc}j_1&j_9&j_2\\-o_b&w_b&p_b\end{array}\right)
\left(\begin{array}{ccc}j_2&j_3&j_{10}\\p_c&q_c&-x_c\end{array}\right)
\left(\begin{array}{ccc}j_4&j_3&j_{11}\\-r_d&q_d&y_d\end{array}\right)
\]
\[
\left(\begin{array}{ccc}j_{12}&j_8&j_5\\-z_e&v_e&s_e\end{array}\right)
\left(\begin{array}{ccc}j_6&j_5&j_9\\-t_f&s_f&w_f\end{array}\right)
\left(\begin{array}{ccc}j_6&j_{10}&j_7\\t_g&x_g&u_g\end{array}\right)
\left(\begin{array}{ccc}j_7&j_{11}&j_8\\-u_h&y_h&v_h\end{array}\right)
\]
\[
\times
\left\{\begin{array}{ccc}j_7&j_{11}&j_8\\J_E&J_F&J_D\end{array}\right\}
\left\{\begin{array}{ccc}j_6&j_{10}&j_7\\J_D&J_F&J_C\end{array}\right\}
\left\{\begin{array}{ccc}j_6&j_5&j_9\\J_B&J_C&J_F\end{array}\right\}
\left\{\begin{array}{ccc}j_{12}&j_8&j_5\\J_F&J_B&J_E\end{array}\right\}
\]
\beq
\left\{\begin{array}{ccc}j_4&j_3&j_{11}\\J_D&J_E&J_A\end{array}\right\}
\left\{\begin{array}{ccc}j_2&j_3&j_{10}\\J_D&J_C&J_A\end{array}\right\}
\left\{\begin{array}{ccc}j_1&j_9&j_2\\J_C&J_A&J_B\end{array}\right\}
\left\{\begin{array}{ccc}j_1&j_4&j_{12}\\J_E&J_B&J_A\end{array}\right\}.
\la{cube2}\eeq

Here $j_{1\!-\!12}$ are the angular momenta attached to the links
of the cube, $(...)$ are $3jm$- and $\{...\}$ are $6j$-symbols. We see
that there is a $6j$ symbol attached to each corner of the even
cube; its arguments are three plaquette momenta $J$ and three link
momenta $j$ intersecting in a given corner. The $3jm$ symbols involve
only link variables $j$.

We have, thus, rewritten all twelve pairs of $D^{J}$-functions entering
a cube as a product of single $D^{j}$-functions, where $j$'s are the
new momenta associated with links. It is understood that this procedure
should be applied to all even cubes of the lattice.  After that,
one has only two $D^{j}$-functions of the same link variable $U$, for
all links of the lattice. It becomes, therefore, straightforward to
integrate over link variables, using \eq{d2link}.

\begin{figure}
\begin{picture}(100,200)(0,0)
\epsfxsize 11.0cm
\epsfysize 11.0cm
\put(60,-30){\epsfbox{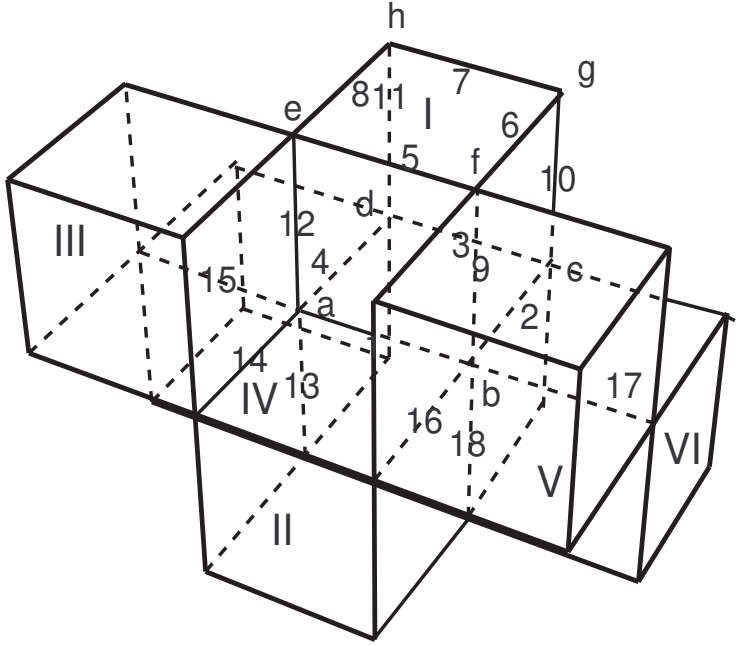} }
\end{picture}
\caption[]{Several cubes combine to produce $6j$ symbols composed of
link momenta $j$. These are the same cubes as in Fig.1.}
\end{figure}

It is convenient to integrate simultaneously over six links entering one
lattice site, because in that way one gets a full contraction over all
`magnetic' numbers.  The derivation is, again, straightforward but
lengthy: the details are given in Appendix C.  The result is that the
$3jm$ factors in \eq{cube2} are contracted with analogous $3jm$ symbols
arising from neighbouring even cubes, and produce $6j$ symbols
attached to every lattice site and composed of the six link momenta $j$
intersecting at a given lattice site. In notations of Fig.3 we get for
the vertices $a,b$:

\beq
{\rm ``a"}\;=
\left\{\begin{array}{ccc}j_1&j_4&j_{12}\\j_{15}&j_{14}&j_{13}
\end{array}\right\},\qquad
{\rm ``b"}\;=
\left\{\begin{array}{ccc}j_1&j_9&j_2\\j_{17}&j_{18}&j_{16}
\end{array}\right\},
\la{veab}\eeq
and similarly for other vertices.  A sign factor $(-1)^{2j}$ should be
attributed to every link of the lattice. As shown in Appendix C it is
actually equivalent to a sign factor $(-1)^{2J}$ attributed to every
lattice plaquette.

\section{Lattice partition function as a product of 6j symbols}

We summarize here the recipe derived in the previous section. One first
divides all 3-cubes into two classes, even and odd ones. They
form a 3-dimensional checker board depicted in Fig.1. All even
cubes are characterized by their plaquette momenta $J$. The edges of
even cubes have link momenta $j$; each link is shared by two
even cubes.

To each of the eight corners of an even cube one attributes a $6j$
symbol of the type

\beq
\left\{\begin{array}{ccc}j_1&j_2&j_3\\J_A&J_B&J_C\end{array}\right\}
\la{jjjJJJ}\eeq
where $J$'s are plaquette and $j$'s are link momenta intersecting
in a given corner of a cube. The rule is that link 1 is perpendicular
to plaquette $A$, link 2 is perpendicular to plaquette $B$ and link
3 is perpendicular to plaquette $C$. Four triades, $(j_1J_BJ_C),
(j_2J_AJ_C), (j_3J_AJ_B)$ and $(j_1j_2j_3)$ satisfy triangle
inequalities.

To each lattice site one attributes a $6j$ symbol of the type

\beq
\left\{\begin{array}{ccc}j_1&j_2&j_3\\j_4&j_5&j_6\end{array}\right\}
\la{jjjjjj}\eeq
where $j$'s are the six link momenta entering a given lattice site.
The rule is that link 4 is a continuation of link 1 lying in the same
direction, link 5 is a continuation of link 2 and link 6 is a
continuation of link 3. Four triades, $(j_1j_2j_3),
(j_1j_5j_6), (j_2j_4j_6)$ and $(j_3j_4j_5)$ satisfy triangle
inequalities.

Actually, each lattice site has {\bf five} $6j$ symbols ascribed to it:
four are originating from the corners of the even cubes adjacent
to the site and are of the type \ur{jjjJJJ}, and one is of the
type \ur{jjjjjj}.

The lattice partition function \ur{Z1} or \ur{Z3} can be {\em
identically} rewritten as a product of the $6j$ symbols described
above. Independent summation over all possible plaquette momenta $J$
and all possible link momenta $j$ is understood. We write the partition
function in a symbolic form:

\[
{\cal Z}=\left[\frac{2}{\beta}I_1(\beta)\right]^{\#{\rm\;of\;plaquettes}}
\sum_{J_P,\;j_l}\quad\prod_{{\rm plaquettes}}
(2J_P+1)\;T_{J_P}(\beta)\;(-1)^{2J_P}\prod_{{\rm links}} (2j_l+1)
\]
\beq
\times \prod_
{{\rm even\;cubes\;corners}}
\left\{\begin{array}{ccc}j&j&j\\J&J&J\end{array}\right\}
\prod_{{\rm lattice\; sites}}
\left\{\begin{array}{ccc}j&j&j\\j&j&j\end{array}\right\}.
\la{Z4}\eeq
The plaquette weights $T_J(\beta)$ are given by \eq{T_J}.
Apart from the sign factor
essentially the same expression was given in refs.
\cite{ACS,HS} \footnote{We are grateful to P.Pobylitsa who has
independently derived \eq{Z4}.}. The sign factor is equal to $\pm 1$ if
the total number of half-integer plaquettes $J$'s is even (odd). Since
plaquettes with half-integer momenta form closed surfaces it may seem
that the sign factor can be omitted. In a general case, however, when
one consideres vacuum averages of operators this is not so, therefore,
it is preferable to keep the sign factor.

\section{Simple example: $d=2$ Yang--Mills}

In a simple exactly soluble case of the 2-dimensional $SU(2)$ theory
every link is shared by only two plaquettes. Therefore, the link
integration is of the type given by \eq{d2link}: it requires that all
plaquettes on the lattice have identical momenta $J$. The partition
function thus becomes a single sum over the common $J$:

\beq
{\cal Z}=\left[\frac{2}{\beta}I_1(\beta)\right]^{\#{\rm\;of\;plaquettes}}
\sum_J\left[T_J(\beta)\right]^{\#{\rm\;of\;plaquettes}},
\la{Zd2}\eeq
the number of plaquettes being equal to $V/a^2$ where $V$ is the
full lattice volume (full area in this case) and $a$ is the lattice
spacing.

A slightly less trivial exercise is to compute the average of the
Wilson loop. Let the Wilson loop be in the
representation $j_s$. It means that one inserts $D^{j_s}(U)$ for all
links along the loop. One gets therefore integrals of two
$D$-functions outside and inside the loop, and integrals of three
$D$-functions for links along the loop. The first integral says that
all plaquettes outside the loop are equal to a common $J$. The second
integral says that all plaquettes inside the loop are equal to a common
$J^\prime$. Integrals along the loop require that $J,J^\prime$ and
$j_s$ satisfy the triangle inequality. We have thus for the average of
the Wilson loop of area $S$:

\beq
\langle W_{j_s}(S)\rangle
=\frac{\sum_J\left[T_J(\beta)\right]^{\frac{V}{a^2}}
\sum_{J^\prime=|J-j_s|}^{J+j_s}\left[T_{J^\prime}(\beta)\,/\,
T_J(\beta)\right]^{\frac{S}{a^2}}}
{\sum_J\left[T_J(\beta)\right]^{\frac{V}{a^2}}}.
\la{avWd2}\eeq
This is an exact expression for the lattice Wilson loop, however we
wish to explore its continuum limit. It implies that $V/a^2\to\infty,
\;S/a^2\to\infty\;$ but $\;S\ll V$; $\beta\to\infty,\; a\to 0\;$
but $\;\beta a^2=4/g_2^2$ fixed, where $g_2^2$ is the physical coupling
constant having the dimension of ${\rm mass}^2$, see \eq{betagen}.

We take $V/a^2\to\infty$ first of all, which requires that only the
$J=0$ term contributes to the sum, with $T_0(\beta)\equiv 1$;
consequently all momenta inside the loop are that of the source,
$J^\prime=j_s$. Taking into account the asymptotics \ur{T_J} of
$T_J(\beta)$ at large $\beta$ we obtain

\beq
\langle W_{j_s}(S)\rangle
=\left[T_{j_s}(\beta)\right]^{\frac{S}{a^2}}
= \exp\left[-\frac{g_2^2}{2}j_s(j_s+1)\,S\right]
\la{avWd2as}\eeq
which is, of course, the well-known area behaviour of the Wilson loop
with the string tension proportional to the Casimir eigenvalue.

\begin{figure}
\begin{picture}(100,150)(0,0)
\epsfxsize 14.0cm
\epsfysize 10cm
\put(50,-45){\epsfbox{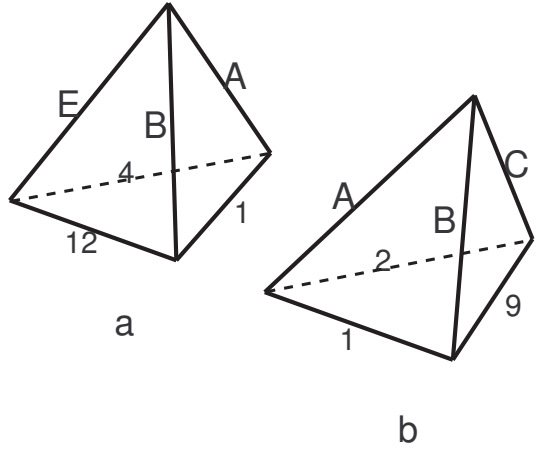} }
\end{picture}
\caption[]{Tetrahedra corresponding to the $6j$ symbols sitting
at vertices $a$ and $b$.}
\end{figure}

\section{Dual lattice: tetrahedra and octahedra}

We now turn to the construction of the dual lattice.

Each $6j$ symbol of the exact partition function \ur{Z4} encodes four
triangle inequalities between the plaquette $J$'s and the link $j$'s.
It is therefore natural to represent all $6j$ symbols by {\bf
tetrahedra} whose six edges have lengths equal to the six momenta of
a given $6j$ symbol.  Four faces of a tetrahedron form four
triangles, so that the triangle inequalities for the momenta are
satisfied automatically.

Let us first consider the eight $6j$ symbols corresponding to the
eight corners of an even cube. These eight $6j$ symbols are given
explicitly in \eq{cube2} with notations shown in Fig.2. Let us
represent all of them by tetrahedra of appropriate edge lengths.
For example, the tetrahedra corresponding to the corners $a$ and $b$
are shown in Fig.4. We denote the plaquette momenta $J_A,...$ just by
their Latin labels $A,B,...$ and the link momenta $j_1,j_2,...$ by
their numerical indices $1,2,..$. We notice immediately that the two
tetrahedra have a pair of equal faces, in this case it is the
triangle $(A,B,1)$. Therefore, we can glue the two tetrahedra
together so that this triangle becomes their common face. The gluing
can be done in two ways. To be systematic we shall always glue
tetrahedra so that their volumes do not overlap.

\begin{figure}
\centerline{\epsfxsize 14.0cm
\epsfysize 8 cm
\epsffile{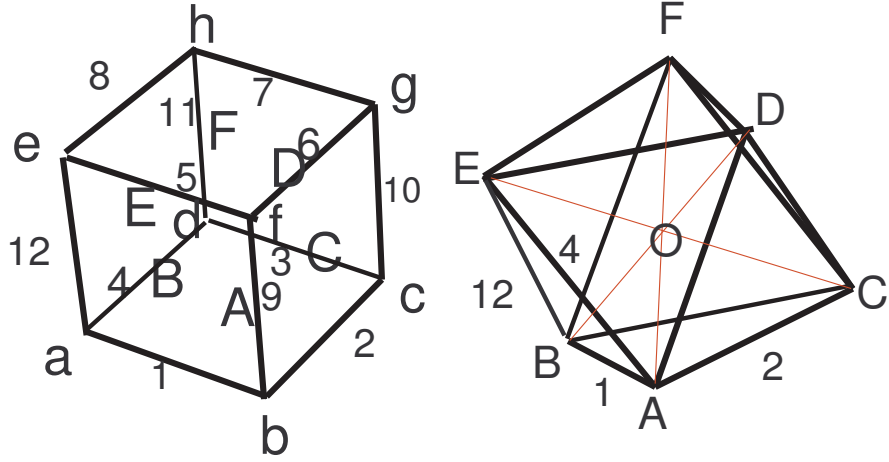}}
\caption[]{Octahedron dual to the even cube.}
\end{figure}

In the same way we glue together other tetrahedra. Being glued
together the eight tetrahedra of the cube form an {\bf octahedron}
shown in Fig.5. Its center point $O$ is connected with six lines
to the vertices denoted as $A\!-\!F$; the lengths of these lines
are equal to the corresponding plaquette momenta $J_{A\!-\!F}$.
The external twelve edges of the octahedron have lengths equal to the
link momenta $j_{1\!-\!12}$. The eight faces of the octahedron
correspond to the eight vertices of the original even cube. One can say
that the octahedron is dual to the cube: the faces become vertices and
{\it vice versa}; the edges remain edges.

It is clear that in a case of generic $J$'s and $j$'s
the octahedron cannot be placed into a flat 3-dimensional space.
Indeed, we have 6+12 $=$ 18 given momenta, that is fixed lengths, but
only 7 points defining the octahedron, including the center one. In
three dimensions that gives 21 d.o.f. from which one has to subtract
3+3 to allow for rigid translations and rotations. Therefore, we are
left with only 15 d.o.f. instead of the needed 18. [In four dimensions
the arithmetic would match: $7\cdot 4\,-\,4\,-\,6\,=\,18$.]

Each even cube of the original lattice has twelve neighbouring
even cubes sharing edges with the first one, and with themselves.
If we represent the neighbour even cubes by their own dual octahedra
those will also share common edges. Does this network of octahedra
cover the space? No, there are holes in between. However, we have
not used yet the $6j$ symbols \ur{jjjjjj} made solely of the link
momenta $j$'s. If we represent these $6j$ symbols by tetrahedra
their triangle faces will coincide with the faces of the octahedra
corresponding to the even cubes adjacent to the site. For example,
if we consider the $6j$ symbols corresponding to the site $a$
(see Fig.3 and \eq{veab}),

\begin{figure}
\centerline{
\epsfxsize7.0cm
\epsffile{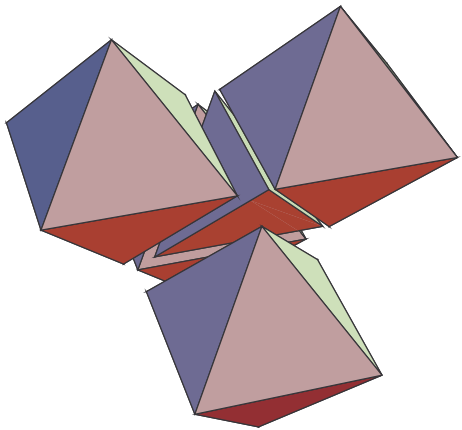}}
\caption[]{A tetrahedron corresponding to the lattice site fits
precisely into a hole between four octahedra corresponding to the
four corners of the even cubes adjacent to the site (shown in motion).}
\end{figure}

\[
\left\{\begin{array}{ccc}j_1&j_4&j_{12}\\j_{15}&j_{14}&j_{13}
\end{array}\right\},
\nonumber\]
it has a common triangle face $(j_1,j_4,j_{12})$ with the octahedron
shown in Fig.5. The other faces of this tetrahedron match the
octahedra dual to the cubes $II,III$ and $IV$, see Fig.3.

Octahedra corresponding to the cubes supplemented by tetrahedra
corresponding to the lattice sites cover the space without holes and
therefore serve as a simplicial triangulation, see Fig.6.

An equivalent view on the dual lattice has been suggested in
ref. \cite{ACS}. One can connect centers of neighbour cubes (both
even and odd) and ascribe plaquette momenta $J$'s to these lines. The
link momenta $j$'s will be then ascribed to diagonal lines connecting
only even neighbour sites of that dual lattice, see Fig.7.

The dual lattice can be understood in two senses. On one hand,
one can build a regular cubic dual lattice with additional face
diagonals like shown in Figs.6 and 7, and ascribe $J$'s and $j$'s to
its edges. On the other hand, since variables living on the links
of the dual lattice are positive numbers, one can build a lattice
{\em with the lengths of edges equal to the appropriate angular momenta}.
We shall always use the dual lattice in this second sense.

\begin{figure}
\begin{picture}(100,150)(0,0)
\epsfxsize7.5cm
\put(85,-20){\epsfbox{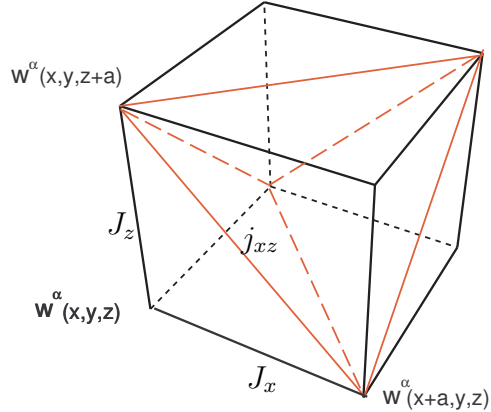} }
\put(137,66){$J_z$}
\put(190,8){$J_x$}
\put(188,61){$j_{xz}$}
\end{picture}
\caption[]{Another view on the dual lattice.}
\end{figure}

\section{Coordinates of the dual lattice as new variables}

In the previous section we have already met with a situation when an
octahedron dual to a cube did not fit into a 3-dimensional
flat space: at least four dimensions were necessary. As one enlarges
the triangulated complex more dimensions are needed to match the
number of degrees of freedom. In the limiting case of an infinite
lattice one needs 6 flat dimensions. This number of dimensions
follows from the number of d.o.f. one has to accommodate: at each
lattice site there are three plaquette momenta $J$ and three link
momenta $j$, and there is a one-to-one correspondence between lattice
sites and the cubes.

Therefore, the dual lattice (understood in the second sense, see above)
spans a 3-dimensional manifold which can be embedded into a
6-dimensional flat space. Notice that it is the maximal number of flat
dimensions needed to embed a general 3-dimensional riemannian manifold;
it can be counted from the number of components of the metric tensor,
which is 6 in three dimensions. Only very special configurations of
$J$'s and $j$'s would be possible to embed into a flat space of less
dimensions.

We are primarily interested in the continuum limit of the lattice
theory, that is in the small $a$, large $\beta$ case. It implies that
large angular momenta $J\sim\sqrt{\beta}$ are involved, and one can
pass from summation over $J$'s and $j$'s to integration over these
variables in the partition function \ur{Z4}. We replace

\beq
\sum_{J=0,1/2,1,...}(2J+1)...\longrightarrow
2\int_0^\infty dJ^2...\;,
\la{sumvint}\eeq
and similarly for the summation over link momenta $j$'s.

The next step is to ascribe a 6-dimensional Lorentz scalar field
$w^\alpha(x),\;\alpha=1,...,6$ to the centers of all cubes of the
original lattice, see Fig.7.
We shall call them coordinates of the dual lattice.
They are scalars because in three dimensions the cubes are scalars.
The argument of the six-component scalar field is the coordinate of
the center of the cube in question, however, we shall consider
$w^\alpha(x)$ as continuous functions.
Since six functions depend only on three coordinates there
are three relations between $w^\alpha(x)$ at any point; these relations
define a curved 3-dimensional manifold whose triangulation is given
by the set of $J$'s and $j$'s.

We next define {\em six-dimensional} angular momenta as differences of
$w^\alpha(x)$ taken at the centers of neighbour cubes:

\bea
J^\alpha_x\left(x+\frac{a}{2},y,z\right)&=&w^\alpha(x+a,y,z)
-w^\alpha(x,y,z)=a\partial_xw^\alpha+\frac{a^2}{2}\partial_x^2w^\alpha
+...\quad,
\nonumber\\
j^\alpha_{xz}\left(x+\frac{a}{2},y,z+\frac{a}{2}\right)
&=&w^\alpha(x+a,y,z)-w^\alpha(x,y,z+a)\!=\!
a(\partial_x-\partial_z)w^\alpha+O(a^2),
\la{defJsix}\eea
and so on. The lengths of these 6-vectors are, by construction, the
lengths of the edges of the dual lattice.

The six functions $w^\alpha(x)$ can be called external coordinates of
the manifold; they induce a metric tensor of the manifold determined
by

\beq
g_{ij}(x)=\partial_iw^\alpha\partial_jw^\alpha.
\la{metric}\eeq
As usual in differential geometry one can define the Christoffel symbol,

\beq
\Gamma_{i,jk}(x)=\frac{1}{2}(\partial_jg_{ik}+\partial_kg_{ij}
-\partial_ig_{jk})=\partial_iw^\alpha\partial_j\partial_kw^\alpha
\equiv (w_i\cdot w_{jk}),
\la{Christ}\eeq
and the Riemann tensor,

\[
R_{ijkl}(x)=\frac{1}{2}(\partial_j\partial_kg_{il}+
\partial_i\partial_lg_{jk}-\partial_j\partial_lg_{ik}-
\partial_i\partial_kg_{jl})+
\Gamma_{m,jk}\Gamma^m_{il}-\Gamma_{m,jl}\Gamma^m_{ik}
\]
\beq
=[(w_{ik}\cdot w_{jl})-g^{pq}(w_p\cdot w_{ik})\,(w_q\cdot w_{jl}]
- [k\leftrightarrow l].
\la{Riemann1}\eeq
The contravariant tensor is inverse to the covariant one,

\beq
g^{ij}g_{jk}=\delta_k^i,
\la{contra}\eeq
and can be used to rise indices, and for contractions. The determinant
of the metric tensor is

\beq
g=\det g_{ij}=\frac{1}{3!}\epsilon^{ijk}\,\epsilon^{lmn}
(w_i\cdot w_l)\,(w_j\cdot w_m)\,(w_k\cdot w_n),
\la{detg}\eeq
and the contravariant metric tensor is

\beq
g^{ij}=\frac{1}{2g}\epsilon^{ikl}\epsilon^{jmn}
(w_k\cdot w_m)\,(w_l\cdot w_n).
\la{contrag}\eeq
There is a useful identity for the antisymmetrized product of two
contravariant tensors, valid in 3 dimensions:

\beq
g^{ik}g^{jl}-g^{il}g^{jk}=\e^{ijm}\,\e^{kln}\,g_{mn}\,/\,g.
\nonumber\eeq
The scalar curvature is obtained as a full contraction:

\[
R=g^{ik}\,g^{jl}\,R_{ijkl}
=\frac{1}{2}(g^{ik}\,g^{jl}-g^{il}\,g^{jk})\,R_{ijkl}
\]
\beq
=\frac{1}{2g^2}\e^{ijk}\e^{i^\prime j^\prime k^\prime}
(w_k\cdot w_{k^\prime})[2g\,(w_{ii^\prime}\cdot w_{jj^\prime})
-\e^{plm}\e^{ql^\prime m^\prime}(w_p\cdot w_{ii^\prime})
(w_q\cdot w_{jj^\prime})(w_l\cdot w_{l^\prime})
(w_m\cdot w_{m^\prime})].
\la{R1}\eeq
Recalling that $w^\alpha$ is a 6-dimensional vector we can rewrite
the scalar curvature in another form:

\beq
R=\frac{1}{72\,g^2}\,\e_{\alpha\beta\gamma\delta\varepsilon\eta}\,
\e_{\alpha^\prime\beta^\prime\gamma^\prime\delta^\prime\varepsilon\eta}\,
\e^{ijk}\,\e^{i^\prime j^\prime k^\prime}\,
\e^{pll^\prime}\e^{qmm^\prime}\;
w_i^\alpha w_{i^\prime}^{\alpha^\prime}
w_j^\beta w_{j^\prime}^{\beta^\prime}
w_k^\gamma w_{k^\prime}^{\gamma^\prime}
w_{lm}^\delta w_{l^\prime m^\prime}^{\delta^\prime}
w_p^\zeta w_q^\zeta.
\la{R2}\eeq
This form makes it clear that the scalar curvature is zero if
$w^\alpha$ has only three nonzero components, which corresponds to a
flat 3-dimensional manifold.

Finally, we would like to point out the Jacobian for the change of
integration variables from the set of the lengths of the tetrahedra
edges, $J_i^2$ and $j_i^2$ given at all lattice sites, to the external
coordinates $w^\alpha$. In the continuum limit this Jacobian is quite
simple. It is given by the determinant of a $6\times 6$ matrix composed
of the second derivatives:

\beq
\prod_x dJ_i^2(x)\, dj_i^2(x) = \prod_x dw^\alpha(x)\,{\rm Jac}(w),\qquad
{\rm Jac}(w)=\det\,w^\alpha_{ij}.
\la{Jac1}\eeq
Since $w^\alpha_{ij}=w^\alpha_{ji}$ there are actually six independent
second derivatives. The Jacobian is zero in the degenerate case when the
triangulation by tetrahedra can be embedded in less than 6 dimensions.

\section{Continuum duality transformation and Bianchi identity}

It is instructive at this point to compare the duality transformation
on the lattice with that in the continuum theory. The continuum
partition function \ur{Z2} can be written with the help of an
additional gaussian integration over the `dual field strength',
$J_{ij}^a$:

\beq
{\cal Z}=\int\! DJ_{ij}^a\;DA_i^a\;\exp\int\!d^3x\left[-\frac{g^2_3}
{4}J_{ij}^2
+\frac{i}{2}J_{ij}^a(\partial_iA_j^a-\partial_jA_i^a+\epsilon^{abc}
A_i^bA_j^c)\right].
\la{Z5}\eeq
\Eq{Z5} is usually called the first-order formalism.

In the Abelian case when the $A_i$ commutator term is absent, integration
over $A_i$ results in the $\delta$-function of the Bianchi identity,

\beq
\partial_iJ_{ij}=0,\qquad{\rm or}\qquad\epsilon_{ijk}\partial_iJ_k=0,
\qquad J_k=\frac{1}{2}\epsilon_{ijk}J_{ij}.
\la{Bianchi1}\eeq
Because of this identity, one can parametrize $J_k=\partial_kw$, and
get for the partition function:

\beq
{\cal Z}_{{\rm abel}}=\int\! Dw\;\exp\int\!d^3x\left[-\frac{g_3^2}{2}
(\partial_kw)^2\right].
\la{Z5A}\eeq
It represents a theory of a free massless scalar field $w$. It is in
accordance with that in a $3d$ Abelian theory there is only one
physical (transverse) polarization. It is easy to check that
gauge-invariant correlation functions of field strengths coincide
with those computed in the original formulation.

In the non-Abelian case integration in $A_i^a$ is more complicated, and
there is no simple Bianchi identity for
$J_k^a=(1/2)\epsilon_{ijk}J_{ij}^a$.  However, one can formally perform
the Gaussian integration over $A_i^a$ \cite{H} resulting in:

\beq
{\cal Z}=\int\!DJ_i^a\;{\rm det}^{\frac{1}{2}}({\cal J}^{-1})\;
\exp\int\!d^3x\left[-\frac{g_3^2}{2}(J_i^a)^2
-\frac{i}{2}\;(\epsilon_{ijm}\partial_jJ_m^a)\;
({\cal J}^{-1})_{ik}^{ab}\;(\epsilon_{kln}\partial_lJ_n^b)\right]
\la{Z5N}\eeq
where ${\cal J}^{-1}$ is the inverse matrix,

\beq
({\cal J}^{-1})_{ik}^{ab}\;\epsilon^{bcd}\epsilon_{klm}J_m^d
=\delta^{ac}\delta_{il};\qquad {\rm det}({\cal J}^{-1})
=({\rm det}\,J_k^a)^{-3}.
\la{Jinv}\eeq
Notice that the second term in the exponent is purely imaginary;
the full partition function is real because for each configuration
$J_i^a(x)$ there exists a configuration with $-J_i^a(x)$, which adds
a complex conjugate expression.

We now turn to the discretized version of the dual theory. As explained
above, we need 6 flat dimensions to embed the dual lattice, and we have
introduced 6-dimensional momenta $J^\alpha$, see \eq{defJsix}. These
momenta apparently satisfy, e.g., the identity (see Fig.7 for
notations):

\[
J^\alpha_z\left(x,y,z+\frac{a}{2}\right)-
J^\alpha_x\left(x+\frac{a}{2},y,z\right)
=w^\alpha(x,y,z+a)-w^\alpha(x+a,y,z)
\]
\beq
=J^\alpha_z\left(x+a,y,z+\frac{a}{2}\right)-
J^\alpha_x\left(x+\frac{a}{2},y,z+a\right),
\la{Bianchi2}\eeq
and similarly for other components. This is nothing but a discretized
version of the Bianchi identity,

\beq
\epsilon_{ijk}\partial_iJ_k^\alpha=0,\qquad \alpha=1,...,6.
\la{Bianchi3}\eeq
Therefore, in 6 dimensions one recovers the simple (flat) form of the
Bianchi identity for the dual field strength. One can say that the
complicated (nonlinear) form of the usual non-Abelian Bianchi identity
is a result of the projection of the flat Bianchi identity onto the
curved colour space.

\section{Wilson loop}

In this section we present the Wilson loop in the representation $j_s$,

\beq
W_{j_s}=\frac{1}{2j_s+1}\Tr\;{\rm P}\,\exp i\oint dx_i A_i^a T^a,
\la{W1}\eeq
in terms of dual variables.

In terms of the original lattice the Wilson loop corresponds to adding
a product of the $D^{j_s}(U)$ functions to all links along the loop,
with a chain contraction of `magnetic' indices. Because of these
insertions, on links containing the loop one has to integrate over
three $D$-functions instead of two as for all other links. As a result
one gets additional $3jm$ symbols along the loop which combine into
the new $9j$ symbols ascribed to all lattice sites, see Appendix D.
For example, the $9j$ symbols ascribed to vertices $a$ and $b$ are
(for notations see Fig.3):

\beq
{\rm ``a"}
=\left\{\begin{array}{ccc}j_4&j_1&j_{12}\\
j_{15}^\prime&{\bf j_s}&j_{15}\\
j_{13}&j_1^\prime&j_{14}\end{array}\right\},\qquad
{\rm ``b"}=
\left\{\begin{array}{ccc}j_2&j_1&j_9\\
j_{17}^\prime&{\bf j_s}&j_{17}\\
j_{18}&j_1^\prime&j_{16}\end{array}\right\}.
\la{9jab}\eeq
The accompanying sign factors are given in Appendix D. Six triades
of the $9j$ symbol, corresponding to all its rows and columns,
satisfy the triangle inequalities.

Contrary to the $6j$ symbol the $9j$ symbol cannot be represented by
a geometrical figure with edges equal to the entries of the
$9j$ symbol. In addition, the link momenta along the loop split now
into pairs: $j_1$ and $j_1^\prime$, $j_{15}$ and $j_{15}^\prime$,
$j_{17}$ and $j_{17}^\prime$, and so on. The `primed' and `non-primed'
angular momenta satisfy triangle inequalities, with the source $j_s$
being the third edge of the triangles. If $j_s$ is an integer, there is
always a contribution with $j_1^\prime=j_1$ (and so on). If $j_s$ is a
half-integer one has necessarily $j_1^\prime\neq j_1$.

Thus, there appears to be a fundamental difference between Wilson loops
in integer and half-integer representations. For integer representations
one can proceed as in the vacuum case and parametrize the dual lattice
sites by the coordinates $w^\alpha(x)$ related to angular momenta through
\eq{defJsix}. In the half-integer case one cannot uniquely parametrize
the dual lattice by the coordinates $w^\alpha(x)$. In the presence of
the Wilson loop in a half-integer representation the dual space
$w^\alpha$ is not simply connected: there is a infinitely thin
cylindrical `hole' in the dual space along the loop.

\section{Asymptotics of the 6j symbols}

In the continuum limit $\beta\to\infty,\quad J,j\to\infty$ one can
replace $6j$ symbols by their asymptotics. The asymptotics was
ingeniously guessed in a seminal paper by Ponzano and Regge \cite{PR}
and later on explicitly derived and improved by Schulten and Gordon
\cite{SG}.  The results of these works can be summarized as follows.

First of all one draws a tetrahedron with edges equal to
$j_n+\frac{1}{2}$, where $j_n$ are the six momenta of a given
$6j$ symbol. It should be stressed that though four momenta triades
satisfy triangle inequalities, the same triades shifted by
$\frac{1}{2}$ need not. In that case the $6j$ symbol is said to
be `classically forbidden', and it is exponentially suppressed at large
$j_n$.

If $j_n$ lie in the `classically allowed' region, the asymptotics is
given by the Ponzano--Regge formula:

\beq
\left\{\begin{array}{ccc}j_1&j_2&j_3\\j_4&j_5&j_6\end{array}\right\}
=\frac{1}{\sqrt{12\pi V(j)}}\cos\left[\sum_n\left(j_n+\frac{1}{2}
\right)\theta_n +\frac{\pi}{4}\right].
\la{PR}\eeq
Here $V(j)$ is the 3-dimensional volume of the tetrahedron and
$\theta_n$ is the dihedral angle in the tetrahedron, corresponding to
the edge $j_n+\frac{1}{2}$. Since we are interested in the large-$j_n$
limit we shall systematically neglect the shifts by $\frac{1}{2}$.
The tetrahedron volume can be found from the Cayley formula:

\beq
V(j)^2=\frac{1}{288}\left|\begin{array}{ccccc}
0&j_4^2&j_5^2&j_6^2&1\\
j_4^2&0&j_3^2&j_2^2&1\\
j_5^2&j_3^2&0&j_1^2&1\\
j_6^2&j_2^2&j_1^2&0&1\\
1&1&1&1&0\end{array}\right|.
\la{Her}\eeq
The dihedral angle corresponding, say, to the edge $j_1$ can be found
from

\beq
\cos\theta_1=\frac{1}{16}\frac{j_1^4+j_1^2(2j_4^2-j_5^2-j_6^2-j_2^2-j_3^2)
+(j_2^2-j_3^2)(j_6^2-j_5^2)}{S(j_1,j_2,j_3)\,S(j_1,j_5,j_6)},
\la{dih1}\eeq
where

\beq
S(j_1,j_2,j_3)=\frac{1}{4}\sqrt{(j_1+j_2+j_3)(j_2+j_3-j_1)
(j_1-j_2+j_3)(j_1+j_2-j_3)}
\la{S}\eeq
is the area of the triangle built on the edges $j_{1,2,3}$. The
dihedral angles are defined such that $0\leq\theta\leq\pi$. Since
in section 6 we have defined 6-dimensional angular momenta
$j^\alpha$ whose lengths are the edges of the tetrahedra, we can
find the dihedral angles from more simple formulae involving scalar
products of momenta in the 6-dimensional space. For example, \eq{dih1}
can be rewritten as

\beq
\cos\theta_1=\frac{(j_1\cdot j_2)(j_1\cdot j_6)-j_1^2(j_2\cdot j_6)}
{\sqrt{j_1^2j_2^2-(j_1\cdot j_2)^2}\;
\sqrt{j_1^2j_6^2-(j_1\cdot j_6)^2}}.
\la{dih2}\eeq
Notice that the angle is defined to be equal to $\pi$ (not $0$!) when
the two vectors, $j_2$ and $j_6$ coincide; it is zero when they point
in the opposite directions. We shall use this formula in what follows.

\section{Angle defect}

The Yang--Mills partition function \ur{Z4} is a product of many $6j$
symbols for each of which we use the asymptotic form \ur{PR} in
approaching the continuum limit. Each {\it cosine} can be written as
a half-sum of exponents of imaginary argument. Therefore, we have to
consider a sum of a product of many imaginary exponents,

\beq
\prod_n^N \cos(\Omega_n)=\frac{1}{2^N}\sum_{\{\epsilon_n=\pm 1\}}
\exp\left(i\sum_n\epsilon_n\Omega_n\right),
\la{prodcos}\eeq
where $\Omega_n$ denotes the argument of the cosine in \eq{PR}, for the
$n^{{\rm th}}$ $6j$ symbol, and one has to sum over all signs
$\epsilon_n=\pm 1$.

The expression in the exponent of \eq{prodcos} can be rearranged as
follows: We first pick one of the edges of the dual lattice,
whose length is a link $j_l$ or a plaquette $J_P$, and combine all
dihedral angles $\theta_n$ related to this edge, as coming from the
$n^{{\rm th}}$ tetrahedron. We then sum over all edges of the dual
lattice.  Therefore, we can write:

\beq
\sum_n\epsilon_n\Omega_n
=\sum_P J_P\left(\sum_{n=1}^4 \epsilon_n\theta_n(J_P)\right)
+\sum_l j_l\left(\sum_{n=1}^6 \epsilon_n\theta_n(j_l)\right),
\qquad \epsilon_n=\pm 1.
\la{angdef1}\eeq
As seen, e.g., from Fig.7, each plaquette $J$ enters four
tetrahedra, therefore the corresponding sum over $n$ in \eq{angdef1}
goes from 1 to 4. Each link $j$ enters six tetrahedra, therefore
in this case the sum is over six dihedral angles $\theta_n(j)$,
with appropriate signs $\epsilon_n$.

Let us consider the contribution to \eq{prodcos} when all signs
$\epsilon_n=+1$, and let us for a moment assume that the dual lattice
spans a 3-dimensional Euclidean manifold.  The sum of the dihedral
angles about an edge is then equal to $4\pi-2\pi=2\pi$ in case of
summing over four tetrahedra, and equal to $6\pi-2\pi=4\pi$ in case of
summing over six tetrahedra.  In the first case we get $\exp(2\pi i
J)=(-1)^{2J}$; in the second case we get $\exp(4\pi i j)=(-1)^{4j}=+1$.
Notice that the sign factor $(-1)^{2J}$ compensates exactly the same
factor in the partition function \ur{Z4}. We conclude that, if the
configuration of the momenta is `flat', there exists a contribution to
the sum \ur{prodcos} that does not oscillate with varying $J$'s and
$j$'s. In fact, there are exactly two such contributions corresponding
to taking {\em all} signs $\epsilon_n=\pm 1$ simultaneously.
Contributions of any other choice of the signs are oscillating fast at
large $J$'s and $j$'s, and thus die out in the continuum limit.

A generic configuration of momenta cannot be embedded into a flat
3-dimensional space, however. Therefore, the sum of dihedral angles
about the edges $J$ and $j$ will, generally, differ from $2\pi$
and $4\pi$, respectively. These differences are sometimes called
angle deficiencies or angle defects (we shall use the second term).
Let us denote them:

\bea
\Theta(J)&=&\sum_{n=1}^4 \theta_n(J)-2\pi,
\la{angledefJ}\\
\Theta(j)&=&\sum_{n=1}^6\theta_n(j)-4\pi.
\la{angledef1}\eea
Our task is to point out contributions to \eq{prodcos} that survive
the continuum limit in a general case when the dual lattice is a
curved 3-dimensional manifold.  To be more precise, we have to consider
the sum of all momenta on the lattice times their angle defects,

\beq
\exp\;i\left[\sum_P J_P\Theta(J_P)+\sum_l j_l\Theta(j_l)\right],
\la{angledefj}\eeq
and to find the contribution of the order of $a^3$ to this exponent,
where $a$ is the lattice spacing. The $O(a^3)$ order is needed to
compensate for the $1/a^3$ factor arising as one goes from summation
over the lattice points to integration over the 3-dimensional space.

In the continuum limit we assume that the momenta are given by the
gradients of a 6-component function $w^\alpha(x)$ having the meaning
of the 6-dimensional coordinates of the dual lattice sites, see
\eq{defJsix}. If we restrict ourselves to the first terms in the
gradient expansion in \eq{defJsix}, the momenta will be
expressed only through three vectors,
$\partial_xw^\alpha,\;\partial_yw^\alpha$ and $\partial_zw^\alpha$.
Three vectors define a flat 3-dimensional space; therefore, the angle
defects $\Theta$ are zero in the first-derivative approximation. To get
a non-zero angle defect it is necessary to expand the momenta in
\eq{defJsix} up to the second derivatives of $w^\alpha$. We shall see
that it is also sufficient in three dimensions.

Since the angle defects $\Theta$'s vanish if $j$'s are taken to the
first approximation of the gradient expansion, it means that the
expansion of $\Theta$'s starts from terms linear in the lattice
spacing $a$. According to \eq{defJsix} the expansion of the momenta
also starts from terms linear in $a$.  Therefore, one can expect that
the expansion of the exponent in \eq{angledefj} starts from the
$O(a^2)$ terms. Were that so, the configuration would be too
`ultraviolet' and would not survive the continuum limit. Fortunately,
there appears to be an exact cancellation of all $O(a^2)$ terms in the
sum over several neighbour edges of the dual lattice, so that
the exponent in \eq{angledefj} proves to be finite in the continuum
limit.

We next embark a rather tedious enterprise of calculating the angle
defects about six plaquette $J$'s in a cube (each entering four
tetrahedra), and about twelve link $j$'s being edges of that cube
(each involved in six tetrahedra, see section 5). Unluckily, it
seems that it is the minimal elementary group  that is being repeated
through the lattice. It means that we have to compute as much as
$6\cdot 4+12\cdot 6=96$ dihedral angles, expressing them through
the first and second gradients of the 6-component function $w^\alpha$
using \eqs{defJsix}{dih2}. This formidable calculation has been performed
by heavily exploiting {\sl Mathematica}. The intermediate
results are very lengthy and we do not present them here. However, the
final result is beautiful. From a direct calculation we obtain:

\bea
\exp\;i\left[\sum_P J_P\Theta(J_P)+\sum_l j_l\Theta(j_l)\right]
&=&\exp\;i\sum_{{\rm points}\;x}a^3 \,\frac{1}{2}\sqrt{g(w)}
\;R(w)
\nonumber\\
\nonumber\\
&=&\exp\;\frac{i}{2}\int d^3x\; \sqrt{g(w)}\;R(w),
\la{defanglej2}\eea
where $g$ is the determinant of the induced metric tensor as given
by \eq{detg}, and $R$ is the corresponding scalar curvature given
by \eq{R1}. Actually, we obtain the expression for the l.h.s. of
\eq{defanglej2} in the form of \eq{R1} (written in components, 384
terms!) from where we recognize that we are dealing with the scalar
curvature.

In fact this result is a concrete realization of a more general theory
developed many years ago by Regge \cite{PR,R}. In these papers it was
shown that the l.h.s. of \eq{defanglej2} should be equal to its
r.h.s. for any simplicial triangulation, {\em provided} it has a smooth
continuum limit. No relation of the scalar curvature $R$ to any concrete
triangulation was given, though. We feel that it is the first time that
this ingenious relation has been derived explicitly for a concrete
triangulation, and the continuum limit shown to exist.

\section{Full partition function}

Having dealt with the $6j$ symbols of the partition function \ur{Z4}
we now turn to the weight factors $T_J(\beta)$. According to \eq{T_J}
at large $\beta$ and $J$ we have:

\[
\prod_{{\rm plaquettes}}T_J(\beta)=
\exp\left[-\sum_{{\rm plaquettes}}\frac{2J^2}{\beta}\right]
\]
\beq
=\exp\left[-\int\! \frac{d^3x}{a^3}\,\frac{2\,J_i^2}{\beta}\right]
=\exp\left[-\int\!d^3x\,\frac{g_3^2}{2}\,
\partial_iw^\alpha\partial_iw^\alpha+O(a^2)\right],
\la{glchlen}\eeq
where the relation \ur{beta} between $\beta$ and the physical coupling
constant $g_3^2$ has been used, together with the gradient expansion
for the angular momenta \ur{defJsix}. Combining \eqs{defanglej2}{glchlen}
and using $\partial_iw^\alpha\partial_iw^\alpha=g_{ii}(w)$ we get finally
for the Yang--Mills partition function:

\beq
{\cal Z}=\int Dw^\alpha(x)\,{\rm Jac}(w)\,g(w)^{-\frac{5}{4}}\,
\exp\int\!d^3x\,\left[-\frac{g_3^2}{2}\, g_{ii}+\frac{i}{2}
\sqrt{g}\;R\right].
\la{Z6}\eeq

The second term is the Einstein--Hilbert action with a purely imaginary
Newton constant; it is invariant under global 6-dimensional rotations of
the external coordinates $w^\alpha(x)$ and, more important, under local
3-dimensional diffeomorphisms $w^\alpha(x)\to w^\alpha(x^\prime(x))$.

The first term in \eq{Z6} can be viewed as a `matter' source,

\beq
-\frac{g_3^2}{2}\int\!d^3x\,g_{ii}=
-\frac{g_3^2}{2}\int\!d^3x\,\sqrt{g}\,T^{ij}g_{ij},
\la{matter1}\eeq
with the stress-energy tensor $T^{ij}\surd{g}=\delta^{ij}$ violating the
invariance under diffeomorphisms. Since it is homogeneous in space it
can be called the `ether'.

The functional measure in \eq{Z6} arises from two sources. One factor
is the Jacobian for the change of variables from the tetrahedra edges
$J$'s and $j$'s to $w^\alpha$, see \eq{Jac1}. The other factor arises
from the tetrahedra volumes in the asymptotics of the $6j$ symbols
\ur{PR}. In the continuum limit the tetrahedron volume can be written
as $V(j)\sim \surd{g}$, and there are 5 tetrahedra per lattice site,
see section 3.

Once the partition function is written in covariant terms one can forget
the origin of the external coordinates $w^\alpha$ (as the coordinates
of the dual lattice) and consider the metric tensor $g_{ij}$ as
independent dynamical variables over which one integrates in \eq{Z6}.
The Jacobian for this change of variables can be easily worked out:
in fact it is the inverse of ${\rm Jac}(w)$ introduced in \eq{Jac1}.
As a result we get the integration measure for the partition function
\ur{Z6}:

\beq
\int Dg_{ij}\,g^{-\frac{5}{4}},\qquad {\rm instead\;\, of}\qquad
\int Dg_{ij}\,g^{-2},
\la{intg}\eeq
which would be the invariant measure in $3d$. We shall get an
independent check of the power $-\frac{5}{4}$ in the next
section. However, it is anyhow a local counterterm not affecting the
physics.

We stress that the partition function written in terms of the metric
tensor {\bf does not contain explicit colour degrees of freedom}.
Nevertheless, implicitly the theory does contain three gluons at short
distances.

Indeed, let us make a simple dimensional analysis of \eq{Z6}. The dimension
of the first term in \eq{Z6} is $g_3^2\partial^2w^2$ (we are just counting
the number of derivatives and the overall power of $w$); the dimension
of the second term is $\partial^3 w^1$.  At short distances where
quantum fluctuations of $w^\alpha(x)$ vary fast, the second term
dominates the first one. Meanwhile, the second term is a fast-oscillating
functional at nonzero $R$. Therefore, the leading contribution to the
functional integral arises from zero-curvature fluctuations of $w^\alpha$,
that is essentially from the 3-dimensional $w^\alpha$. Being plugged
into the first term, the three components of $w^\alpha$ describe three
massless scalar fields. These fields correspond to three gluons of
$SU(2)$ with one physical (transverse) polarization. It should be
paralleled to \eq{Z5A} for free electrodynamics. This is the correct
result for the non-Abelian theory at short distances in three dimensions.

At large distances or at low field momenta the dominant term is,
on the contrary, the first one as it has less derivatives. It describes
{\em six} (instead of three) {\em massless scalar} degrees of freedom.
It is the correct number of {\em gauge-invariant} degrees of freedom in
the $SU(2)$ theory. However, the theory remains strongly nonlinear,
and it is not clear so far whether massless modes survive in the
physical spectrum.

\section{Quantum gravity from first-order continuum formalism}

In this section we give another derivation of the partition function
\ur{Z6} directly in the continuum theory starting from the first-order
formalism, see section 7. We shall show that the two terms in the
exponent of \eq{Z5} are in fact in one-to-one correspondence with the two
terms in \eq{Z6}, and that the integration measure coincides with that of
\eq{intg}.

Actually, it has been already derived in the previous section that the
first terms of eqs.~(\ref{Z5}) and (\ref{Z6}) are equal:

\beq
S_1=-\frac{g_3^2}{2}\int\!d^3x\;(J_i^a)^2
=-\frac{g_3^2}{2}\int\!d^3x\;\partial_iw^\alpha\partial_iw^\alpha
=-\frac{g_3^2}{2}\int\!d^3x\;g_{ii}.
\la{1st}\eeq
Let us derive a less trivial relation for the second terms:

\beq
S_2=\frac{i}{2}\int\!d^3x\;\;\epsilon^{ijk}\,J_i^a
(\partial_jA_k^a-\partial_kA_j^a+\epsilon_{abc}\,A_j^bA_k^c)
=\frac{i}{2}\int\!d^3x\;\sqrt{g}R.
\la{2nd}\eeq
This derivation will be done in two steps. We shall first show,
following Witten \cite{Wit}, that the l.h.s. of \eq{2nd} can be presented
as a certain Chern--Simons term. Second, we shall show that it is
formally equal to the Einstein--Hilbert action. A subtle question about
the integration measure will be discussed at the end of the section.

The l.h.s. of \eq{2nd} is apparently invariant under ordinary gauge
transformations:

\beq
\delta A_i^a=-\partial_i\delta^{ab}\omega^b+\e_{abc}\,\omega^b A^c_i
=-D_i^{ab}(A)\omega^b, \quad
\delta J_i^a=\e_{abc}\,\omega^b J^c_i,
\la{gt1}\eeq
where $D_i^{ab}(A)=\partial_i\delta^{ab}+\e_{acb}\,A_i^c$ is the covariant
derivative.

Less evident, it is also invariant under the following local
transformation:

\beq
\delta J_i^a=-\partial_i \rho^a-\e_{abc}\,\rho^b A^c_i,\qquad
\delta A_i^a=0.
\la{gt2}\eeq
Indeed, after integrating by parts we obtain the following variation
of the action:

\beq
\delta S_2=\frac{i}{2}\int\! d^3 x\, \rho^a\e_{ijk}\,D^{ab}_i(A)
F^b_{jk},\qquad F^b_{jk}=
\partial_jA_k^b-\partial_kA_j^b+\epsilon_{bcd}\,A_j^c A_k^d.
\la{var}\eeq
This variation is zero owing to the Bianchi identity,
$\e_{ijk}\,D^{ab}_i F^b_{jk}=0$.

The two transformations combined form a 6-parameter gauged Poincare
group, called $ISO(3)$. Indeed, let us introduce three `momenta'
generators $P_i$ and three `angular momenta' generators $L_i$ satisfying
the Poincare algebra,

\beq
[P_a\,P_b]=0,\qquad [L_a\,L_b]=i\e_{abc}\,L_c,\qquad
[L_a\,P_b]=i\e_{abc}\,P_c.
\la{Poinc}\eeq
We next introduce a 6-component vector field $\hat{B}_i$:

\beq
\hat{B}_i=J^a_i\,P_a+A_i^a\,L_a\equiv B_i^\alpha T^\alpha,\qquad
T^\alpha=\left\{\begin{array}{cc}P_a,&\quad\alpha=a=1,2,3,\\
L_a,&\quad\alpha=3+a=4,5,6.\end{array}\right.
\la{Bhat}\eeq
Its gauge transformation has the standard form:

\beq
\hat{B}_i\rightarrow S^{-1} \hat{B}_i S+iS^{-1}\partial_i S,
\qquad S=\exp[i\rho^aP_a+i\omega^aL_a].
\la{gt3}\eeq
Using the Poincare algebra \ur{Poinc} it is easy to check that its
infinitesimal form coincides with \eqs{gt1}{gt2}.

Since the l.h.s. of \eq{2nd} is invariant under these 6-parameter
transformations, it can be rewritten in an explicitly $ISO(3)$-invariant
form. To that end we notice that the invariant tensor of this group is

\beq
M_{\alpha\beta}=\left(\begin{array}{cc}0&1\\1&0\end{array}\right),
\la{M}\eeq
where ``1'' is a unit $3\times 3$ matrix. This matrix defines a
scalar product, $B^\alpha M^{\alpha\beta} C^\beta$, which is invariant
under global ($x$-independent) transformations \ur{gt3}. With the help
of this invariant tensor we build a {\em local} gauge-invariant action
having the form of the Chern--Simons term:

\beq
S_2=\frac{i}{2}\int\!d^3x\;\epsilon^{ijk}\,M_{\alpha\beta}\,
B^\alpha_i\left(\partial_jB_k^\beta+\frac{1}{3}F^\beta_{\gamma\delta}
B_j^\gamma B_k^\delta\right),
\la{ChS}\eeq
where $F^\alpha_{\beta\gamma}=-F^\alpha_{\gamma\beta}$ are the $ISO(3)$
structure constants. Explicitly,

\[
F^\alpha_{bc}=0,\qquad F^a_{3+b,c}=\epsilon_{abc}
\]
\beq
F^{3+a}_{3+b,c}=F^a_{3+b,3+c}=0,\qquad F^{3+a}_{3+b,3+c}=\epsilon_{abc}.
\la{fabc}\eeq
Using the definition \ur{Bhat} it is easy to check that \eq{ChS}
coincides with the l.h.s. of \eq{2nd}, however it is explicitly invariant
under the 6-parameter gauge transformation \ur{gt3}.

\Eq{ChS} has the form of the Chern--Simons term in a Yang--Mills
theory. Though our derivation above is for the gauge group $SU(2)$
it is trivially generalized to any Lie group: to that end it is
sufficient to replace the $SU(2)$ structure constants $\e_{abc}$ by
the structure constants $f_{abc}$ of the gauge group under consideration.
We also note in passing that in {\em four} dimensions the
mixed $iJ_{\mu\nu}^aF_{\mu\nu}^a(A)$ term of the first-order formalism
also posseses an additional local symmetry. To unveil it, it is
sufficient to replace the scalar parameter $\rho^a$ in the transformation
\ur{gt2} by a 4-vector parameter $\rho_\mu^a$:  the invariance is again
due to the Bianchi identity, this time in four dimensions.

The second step in the derivation is more standard. Introducing the
{\em dreibein} $e_i^a,\quad a=1,2,3,$ satisfying the condition
$e^a_ie^{bi}=\delta^{ab}$, so that the metric tensor is
$g_{ij}=e^a_ie^a_j$, and the {\em connection}

\beq
\omega^{ab}_i=\frac{1}{2}e^{ak}(\partial_ie_k^b-\partial_ke_i^b)
-\frac{1}{2}e^{bk}(\partial_ie_k^a-\partial_ke_i^a)
-\frac{1}{2}e^{ak}e^{bl}e^c_i(\partial_ke_l^c-\partial_le_k^c),
\la{conn}\eeq
one can identically rewrite $\surd{g}R$ as

\beq
\sqrt{g}R=\frac{1}{2}\e^{ijk}\,e^a_i\,(\partial_j\omega^a_k
-\partial_k\omega^a_j+\e_{abc}\,\omega^b_j\omega^c_k)
\la{Rg1}\eeq
where $\omega_{ai}=\omega^a_i=\frac{1}{2}\e_{abc}\omega^{bc}_i$.
Finally, one notices that, if one makes an identification of the
dreibein with the dual field strength, $e^a_i=J^a_i$, and of the
connection with the Yang--Mills potential, $\omega_i^a=A^a_i$,
then \eq{Rg1} takes exactly the form of the l.h.s. of \eq{2nd}.
This parallel has been first noticed in ref. \cite{Lun1}.

There is a subtle point in this formal derivation, however. The
use of the first-order formalism implies that one integrates both over
$J_i^a$ and over $A_i^a$ (see \eq{Z5}) or, equivalently, over the
dreibein and over the connection, independently. Meanwhile, the use of
the Einstein quantum gravity implies that the connection is
rigidly related to the dreibein via \eq{conn}, moreover, we have
explicitly used this relation in the above derivation. In ref. \cite{Wit}
Witten has presented arguments that one can, nevertheless,
integrate over the connection as independent variable. However, the
arguments rely upon the use of the equations of motion (one of which
is the relation \ur{conn}), and that might be dangerous in full
quantum field theory.

The present paper gives a different kind of argument that the two
approaches are in fact equivalent. We start with the Yang--Mills
partition function. On one hand it can be presented in the first-order
formalism where one integrates independently over $J_i^a$ (the dreibein)
and over $A_i^a$ (the connection). On the other hand we have shown that
the Yang--Mills theory is equivalent to quantum gravity where one
integrates over the external coordinates $w^\alpha$, or over the metric
tensor, or over the dreibein {\em only}.

Since pure gravity can be rewritten as a Chern--Simons term \ur{ChS}, it
is actually a topological field theory \cite{Wit}, with no real
propagating particles. It is the `ether' term that violates the
invariance under diffeomorphisms and restores the propagation of gluons,
as it should be in the Yang--Mills theory, see the end of the previous
section.

Finally, we would like to remark that the integration measure \ur{intg}
could be anticipated from the first-order formalism as well. Indeed,
integrating in \eq{Z5N} over $A_i^a$ one gets \eq{Jinv}, where the
integration measure over the dreibein is
$(\det\,J)^{-3/2}\sim g^{-3/4}$. The Jacobian for the change of variables
from the dreibein to the metric tensor is $de_i^a\sim dg_{ij}\,g^{-1/2}$.
Adding the powers we obtain: $-\frac{3}{4}-\frac{1}{2}=-\frac{5}{4}$, as
in \eq{intg}.

\section{Conclusions and outlook}

In this paper, we have studied the dual transformation of the $SU(2)$
Yang--Mills theory in 3 dimensions, both from the continuum and
lattice points of view.

On the lattice, one can introduce dual variables being the angular
momenta of the plaquettes ($J$'s) supplemented by those associated with
the links ($j$'s). The partition function can be identically rewritten
as a product of $6j$ symbols made of those angular momenta. A Wilson loop
corresponds to taking a product of $9j$ symbols replacing the $6j$
symbols along the loop. One can construct a dual lattice made of
tetrahedra whose edges have the lengths equal to $J$'s and $j$'s; the
tetrahedra span a $3d$ curved manifold which can be embedded into a
flat $6d$ space.

In the continuum limit the angular momenta are large, and we have
introduced continuum $6d$ Euclidean external coordinates $w^\alpha(x)$ to
describe the curved dual space. The Bianchi condition for the Yang--Mills
field strength has been shown to be trivially soluble in flat six
dimensions.

At large angular momenta one can use the asymptotics of the $6j$ symbols,
given by Ponzano and Regge. Using a specific simplicial triangulation
of the dual space (as dictated by the original lattice) we have shown
that the product of the $6j$ symbols does have a smooth continuum limit
which appears to be the Einstein--Hilbert action, with the metric tensor
$g_{ij}$ and the scalar curvature $R$ expressed through the flat external
coordinates $w^\alpha(x)$. This result cannot be considered as
particularly new (it is the cornerstone of the Regge's simplicial
gravity), however, to our best knowledge it is the first time that the
result has been explicitly derived from a concrete triangulation of the
curved space, and the continuum limit shown to exist. We have also
found the integration measure  for the continuum limit.

The continuum Yang--Mills partition function can be rewritten as
a quantum gravity theory but with an `ether' term violating the
invariance in respect to general coordinate transformations or
diffeomorphisms. This term, however, revives gluons at short distances,
in contrast to the topological pure gravity theory where no particles
propagate.

The presentation of the Yang--Mills theory in a quantum gravity form
\ur{Z6} is explicitly colour gauge-invariant since the metric tensor of
the dual space is colour-neutral. We have, thus, formulated the
Yang--Mills theory solely in terms of colourless `glueball' degrees of
freedom \footnote{A somewhat similar line was developed in
ref. \cite{HJLS} for the $3+1$ dimensional Yang--Mills theory in the
Hamiltonian approach; see also ref. \cite{Lun2}.}.
It turns out to be an interacting theory of {\em six massless
scalar fields}. Nevertheless, at small distances it correctly reproduces
the propagation of gluons. It is not clear to us at the moment how
to proceed best in order to reveal its large-distance behaviour. Let us
indicate a few possibilities.

One possibility is to exploit the fact that the pure quantum gravity
theory is topological, therefore essentially a free theory. One can try to
make a perturbative expansion in $g_3^2$ about it.

Another possibility is to make use of the fact that the Chern-Simons term
\ur{ChS} can be obtained from integrating over heavy fermions, in this
case belonging to some $ISO(3)$ representation. The subsequent
integration over bosonic fields $A_i,\,J_i$ is trivial since there is
no kinetic energy term for those fields: the result would be a local
four-fermion theory with infinitely heavy fermions; it might be soluble,
at least in the large-$N_c$ limit.

Probably the most promising possibility is to pursue the analogy with
and methods of quantum gravity. One can average \eq{Z6} over $3d$
diffeomorphisms: the second term is invariant, the first term is not.
Integrating the first term over diffeomorphisms produces
diffeomorphism-invariant effective action containing growing
powers in the curvature. The effective action may lead to a nonzero
v.e.v. of the scalar curvature, and that may yield a mass gap for the
diffeomorphism-non-invariant correlation functions, like the correlation
functions of $F_{\mu\nu}^2$.

There are several other tasks for the future, lying on the surface. First,
it would be interesting to generalize the present approach to colour
groups other than $SU(2)$. In view of the sad fact that the theory of the
``$6j$ symbols'' for higher Lie groups is not too developed it will be
probably difficult to make a straightforward generalization of the
lattice formulation. A more promising approach would be to start from
the first-order formalism, the more so that the wide local symmetry
revealed in section 12 can be directly generalized to any Lie group.
Second, it would be interesting to make a transformation similar to that
of this paper in $d=4$. The lattice $6j$ symbols have been known for a
while in this case \cite{HS} (for the SU(2) colour), however it again
seems that the first-order formalism is a more promising start, due to
the additional gauge symmetry noticed in section 12.
\vskip .5true cm

We are grateful to Pavel Pobylitsa for many fruitful discussions.
D.D. acknowledges a very useful conversation with Ben Mottelson.
V.P. is grateful to NORDITA for the hospitality extended to him in
Copenhagen, in particular for a partial support by a Nordic Project
grant. The work was supported in part by the Russian Foundation for
Basic Research grant 97-27-15L.

\vskip 1true cm

\section*{Appendix A. D-functions, 3jm, 6j and 9j symbols}

Wigner $D$-functions are eigenfunctions of the square of the angular
momentum operator (written in terms of, say, three Euler angles
$\alpha,\,\beta,\,\gamma$),

\beq
{\bf J}^2 D^J_{mn}(\alpha,\beta,\gamma)
=J(J+1)D^J_{mn}(\alpha,\beta,\gamma),\qquad
J=0,\,\frac{1}{2},\,1,\,\frac{3}{2}...,\qquad -J\leq m,n\leq +J,
\la{eigenf}\eeq
and can be said to be eigenfunctions of a spherical top; they are
$(2J+1)^2$-fold degenerate. The `magnetic' quantum numbers $m,\,n$
have the meaning of the projections of the angular momentum of a
spherical top on the third axes in the `body-fixed' and `lab' frames.
One can parametrize a $2\times 2$ unitary matrix by Euler angles as

\beq
U=\exp(i\alpha\tau^3)\exp(i\beta\tau^2)\exp(i\gamma\tau^3).
\la{Euler}\eeq
It is convenient to use the unitary matrix $U$ as a formal argument
of the $D$-functions. Their main properties are:

\begin{itemize}
\item Multiplication law:
\beq
D^J_{kl}(U_1U_2)=D^J_{km}(U_1)D^J_{ml}(U_2) \qquad {\rm (summation\;
over\; repeated\; indices\; understood)}.
\la{multiplication}\eeq

\item Unitarity:
\beq
D^J_{kl}(U^{\dagger})=\left(D^J_{lk}(U)\right)^* \qquad {\rm (``*"\;
denotes\; complex\; conjugate)}.
\la{unitarity}\eeq

\item Phase condition:
\beq
\left(D^J_{lk}(U)\right)^* = (-1)^{l-k}D^J_{-l,-k}(U),\qquad
D^J_{kl}(1)=\delta^{(2J+1)}_{kl}.
\la{phase}\eeq

\item Orthogonality and normalization:
\beq
\int dU D^{J_1}_{kl}(U^{\dagger}) D^{J_2}_{mn}(U)
=\frac{1}{2J_1+1}\delta_{J_1J_2}\;\delta_{kn}\;\delta_{lm}.
\la{orthonormal}\eeq
Integration here is over the Haar measure:
\beq
\int dU...=\int d(SU)...=\int d(US)...;\qquad\int dU=1.
\la{Haar}\eeq

\item Completeness (the $\delta$-function is understood in the Haar
measure sense):
\beq
\delta(U,V)=\sum_J(2J+1)D^J_{kl}(U^{\dagger})D^J_{lk}(V).
\la{completeness}\eeq

\item Matrix element:
\beq
\int dU D^{J_1}_{a_1b_1}(U)D^{J_2}_{a_2b_2}(U) D^{J_3}_{a_3b_3}(U)
=\left(\begin{array}{ccc}J_1&J_2&J_3\\a_1&a_2&a_3\end{array}\right)
\left(\begin{array}{ccc}J_1&J_2&J_3\\b_1&b_2&b_3\end{array}\right),
\la{me}\eeq
where $(...)$ denote $3jm$ symbols.

\item Decomposition of a direct product of ireps:
\beq
D^{J_1}_{a_1b_1}(U)D^{J_2}_{a_2b_2}(U)
=\sum_J(2J+1)
\left(\begin{array}{ccc}J&J_1&J_2\\-c&a_1&a_2\end{array}\right)
\left(\begin{array}{ccc}J&J_1&J_2\\-d&b_1&b_2\end{array}\right)
(-1)^{d-c}\;D^J_{cd}(U).
\la{decomposition}\eeq
The last two factors may be replaced by $D^J_{-d,-c}(U^\dagger)$
using \eq{phase}.
\end{itemize}

The $3jm$ symbols are symmetric under cyclic permutations of the
columns. An interchange of two columns gives a sign factor:

\beq
\left(\begin{array}{ccc}j_1&j_2&j_3\\k&l&m\end{array}\right)
=(-1)^{j_1+j_2+j_3}
\left(\begin{array}{ccc}j_2&j_1&j_3\\l&k&m\end{array}\right),
\qquad {\rm etc.}
\la{perm1}\eeq
If one changes the signs of all `magnetic' quantum numbers
or projections, the $3jm$ symbol also gets a sign factor:

\beq
\left(\begin{array}{ccc}j_1&j_2&j_3\\k&l&m\end{array}\right)
=(-1)^{j_1+j_2+j_3}
\left(\begin{array}{ccc}j_1&j_2&j_3\\-k&-l&-m\end{array}\right).
\la{sign1}\eeq

A ``practical'' definition of the $6j$ symbol $\{...\}$ is via a
contraction over projections in three $3jm$ symbols:

\[
\sum_{klm}(-1)^{j_4-k+j_5-l+j_6-m}
\left(\begin{array}{ccc}j_5&j_1&j_6\\l&p&-m\end{array}\right)
\left(\begin{array}{ccc}j_6&j_2&j_4\\m&q&-k\end{array}\right)
\left(\begin{array}{ccc}j_4&j_3&j_5\\k&r&-l\end{array}\right)
\]
\beq
=\left(\begin{array}{ccc}j_1&j_2&j_3\\-p&-q&-r\end{array}\right)
\left\{\begin{array}{ccc}j_1&j_2&j_3\\j_4&j_5&j_6\end{array}\right\}.
\la{6jdef1}\eeq
The summation over projections $k,l,m$ is such that $p=m-l$,
$q=k-m$ and $r=l-k$ are kept fixed.

Another definition of the $6j$ symbol is via the full contraction of
projections in four $3jm$ symbols:

\[
\sum_{klmnop}(-1)^{j_4+n+j_5+o+j_6+p}
\left(\begin{array}{ccc}j_1&j_2&j_3\\k&l&m\end{array}\right)
\left(\begin{array}{ccc}j_1&j_5&j_6\\k&o&-p\end{array}\right)
\left(\begin{array}{ccc}j_4&j_2&j_6\\-n&l&p\end{array}\right)
\left(\begin{array}{ccc}j_4&j_5&j_3\\n&-o&m\end{array}\right)
\]
\beq
=\left\{\begin{array}{ccc}j_1&j_2&j_3\\j_4&j_5&j_6\end{array}\right\}.
\la{6jdef2}\eeq
Since the three $j$'s of any $3jm$ symbol satisfy the triangle
inequalities, e.g. $|j_1-j_2|\leq j_3\leq j_1+j_2$, etc., the following
four triades of the $6j$ symbols have to satisfy the triangle
inequalities: $(j_1j_2j_3)$, $(j_1j_5j_6)$, $(j_2j_4j_6)$ and
$(j_3j_4j_5)$; otherwise, the $6j$ symbol is zero.

The $6j$ symbols are symmetric under permutation of any of two columns
and under interchange of the upper and lower arguments simultaneously
in any two columns, e.g.,

\beq
\left\{\begin{array}{ccc}j_1&j_2&j_3\\j_4&j_5&j_6\end{array}\right\}
=\left\{\begin{array}{ccc}j_1&j_3&j_2\\j_4&j_6&j_5\end{array}\right\}
=\left\{\begin{array}{ccc}j_4&j_2&j_6\\j_1&j_5&j_3\end{array}\right\},
\qquad{\rm etc.}
\la{sym6j}\eeq

A full contraction of six $3jm$ symbols yields the $9j$ symbol:

\[\sum
\left(\begin{array}{ccc}j_1&j_2&j_3\\k&l&m\end{array}\right)
\left(\begin{array}{ccc}j_4&j_5&j_6\\n&o&p\end{array}\right)
\left(\begin{array}{ccc}j_7&j_8&j_9\\q&r&s\end{array}\right)
\left(\begin{array}{ccc}j_1&j_4&j_7\\k&n&q\end{array}\right)
\]
\beq
\times\left(\begin{array}{ccc}j_2&j_5&j_8\\l&o&r\end{array}\right)
\left(\begin{array}{ccc}j_3&j_6&j_9\\m&p&s\end{array}\right)
=\left\{\begin{array}{ccc}j_1&j_2&j_3\\j_4&j_5&j_6\\j_7&j_8&j_9
\end{array}\right\}.
\la{9jdef}\eeq
$9j$ symbol is symmetric under transposition and under even
permutations of rows and columns; under odd permutations it acquires a
sign factor $(-1)^{j_1+...+j_9}$.  As follows from the definition, six
momenta triades corresponding to the rows and columns of the $9j$
symbol satisfy triangle inequalities.

A convenient reference book on $D$-functions, $3jm$, $6j$ and $9j$
symbols is ref.\cite{MVKh} from where we have borrowed the definitions.

\section*{Appendix B. 6j symbols in an `even' cube}

In this Appendix we make the decomposition of two plaquettes
$D^{J}$-functions into a sum of single $D^{j}$-functions labelled by
link angular momenta $j$. Then we assemble the arising $3jm$ symbols
into $6j$ symbols attached to the corners of the even cubes.
The notations are given in Fig.2.

We find it convenient (though not necessary) to write the decomposition
for the pairs containing $U_{1,4,12,6,7,10}$ (these are links sitting
at lower left and upper right corners of the cube) in terms of $D(U)$,
and the rest in terms of $D(U^\dagger)$.

Exploiting \eq{decomposition} of Appendix A we get:

\[
D^{J_A}_{i_ai_b}(U_1)\;D^{J_B}_{j_bj_a}(U_1^\dagger)\]\[
=(-1)^{j_a-j_b}\sum_{j_1}(2j_1+1)
\left(\begin{array}{ccc}j_1&J_A&J_B\\-o_a&i_a&-j_a\end{array}\right)
\left(\begin{array}{ccc}j_1&J_A&J_B\\-o_b&i_b&-j_b\end{array}\right)
(-1)^{o_b-o_a}D^{j_1}_{o_ao_b}(U_1),
\]
\[
D^{J_A}_{i_bi_c}(U_2)\;D^{J_C}_{k_bk_c}(U_2^\dagger)\]\[
=(-1)^{k_b-k_c}\sum_{j_2}(2j_2+1)
\left(\begin{array}{ccc}j_2&J_A&J_C\\-p_b&i_b&-k_b\end{array}\right)
\left(\begin{array}{ccc}j_2&J_A&J_C\\-p_c&i_c&-k_c\end{array}\right)
D^{j_2}_{-p_c,-p_b}(U_2^\dagger),
\]
\[
D^{J_D}_{l_dl_c}(U_3)\;D^{J_A}_{i_ci_d}(U_3^\dagger)\]\[
=(-1)^{i_d-i_c}\sum_{j_3}(2j_3+1)
\left(\begin{array}{ccc}j_3&J_D&J_A\\-q_d&l_d&-i_d\end{array}\right)
\left(\begin{array}{ccc}j_3&J_D&J_A\\-q_c&l_c&-i_c\end{array}\right)
D^{j_3}_{-q_c,-q_d}(U_3^\dagger),
\]
\[
D^{J_E}_{m_am_d}(U_4)\;D^{J_A}_{i_di_a}(U_4^\dagger)\]\[
=(-1)^{i_a-i_d}\sum_{j_4}(2j_4+1)
\left(\begin{array}{ccc}j_4&J_E&J_A\\-r_a&m_a&-i_a\end{array}\right)
\left(\begin{array}{ccc}j_4&J_E&J_A\\-r_d&m_d&-i_d\end{array}\right)
(-1)^{r_d-r_a}D^{j_4}_{r_ar_d}(U_4),
\]
\[
D^{J_B}_{j_ej_f}(U_5)\;D^{J_F}_{n_fn_e}(U_5^\dagger)\]\[
=(-1)^{n_e-n_f}\sum_{j_5}(2j_5+1)
\left(\begin{array}{ccc}j_5&J_B&J_F\\-s_e&j_e&-n_e\end{array}\right)
\left(\begin{array}{ccc}j_5&J_B&J_F\\-s_f&j_f&-n_f\end{array}\right)
D^{j_5}_{-s_f,-s_e}(U_5^\dagger),
\]
\[
D^{J_C}_{k_fk_g}(U_6)\;D^{J_F}_{n_gn_f}(U_6^\dagger)\]\[
=(-1)^{n_f-n_g}\sum_{j_6}(2j_6+1)
\left(\begin{array}{ccc}j_6&J_C&J_F\\-t_f&k_f&-n_f\end{array}\right)
\left(\begin{array}{ccc}j_6&J_C&J_F\\-t_g&k_g&-n_g\end{array}\right)
(-1)^{t_g-t_f}D^{j_6}_{t_ft_g}(U_6),
\]
\[
D^{J_F}_{n_hn_g}(U_7)\;D^{J_D}_{l_gl_h}(U_7^\dagger)\]\[
=(-1)^{l_h-l_g}\sum_{j_7}(2j_7+1)
\left(\begin{array}{ccc}j_7&J_F&J_D\\-u_h&n_h&-l_h\end{array}\right)
\left(\begin{array}{ccc}j_7&J_F&J_D\\-u_g&n_g&-l_g\end{array}\right)
(-1)^{u_g-u_h}D^{j_7}_{u_hu_g}(U_7),
\]
\[
D^{J_F}_{n_en_h}(U_8)\;D^{J_E}_{m_hm_e}(U_8^\dagger)\]\[
=(-1)^{m_e-m_h}\sum_{j_8}(2j_8+1)
\left(\begin{array}{ccc}j_8&J_F&J_E\\-v_e&n_e&-m_e\end{array}\right)
\left(\begin{array}{ccc}j_8&J_F&J_E\\-v_h&n_h&-m_h\end{array}\right)
D^{j_8}_{-v_h,-v_e}(U_8^\dagger),
\]
\[
D^{J_C}_{k_bk_f}(U_9)\;D^{J_B}_{j_fj_b}(U_9^\dagger)\]\[
=(-1)^{j_b-j_f}\sum_{j_9}(2j_9+1)
\left(\begin{array}{ccc}j_9&J_C&J_B\\-w_b&k_b&-j_b\end{array}\right)
\left(\begin{array}{ccc}j_9&J_C&J_B\\-w_f&k_f&-j_f\end{array}\right)
D^{j_9}_{-w_f,-w_b}(U_9^\dagger),
\]
\[
D^{J_D}_{l_cl_g}(U_{10})\;D^{J_C}_{k_gk_c}(U_{10}^\dagger)\]\[
=(-1)^{k_c-k_g}\sum_{j_{10}}(2j_{10}+1)
\left(\begin{array}{ccc}j_{10}&J_D&J_C\\-x_c&l_c&-k_c\end{array}\right)
\left(\begin{array}{ccc}j_{10}&J_D&J_C\\-x_g&l_g&-k_g\end{array}\right)
(-1)^{x_g-x_c}D^{j_{10}}_{x_cx_g}(U_{10}),
\]
\[
D^{J_E}_{m_dm_h}(U_{11})\;D^{J_D}_{l_hl_d}(U_{11}^\dagger)\]\[
=(-1)^{l_d-l_h}\sum_{j_{11}}(2j_{11}+1)
\left(\begin{array}{ccc}j_{11}&J_E&J_D\\-y_d&m_d&-l_d\end{array}\right)
\left(\begin{array}{ccc}j_{11}&J_E&J_D\\-y_h&m_h&-l_h\end{array}\right)
D^{j_{11}}_{-y_h,-y_d}(U_{11}),
\]
\beq
D^{J_B}_{j_aj_e}(U_{12})\;D^{J_E}_{m_em_a}(U_{12}^\dagger)\]\[
=(-1)^{m_a-m_e}\!\sum_{j_{12}}(2j_{12}+1)
\left(\begin{array}{ccc}j_{12}&J_B&J_E\\-z_a&j_a&-m_a\end{array}\right)
\left(\begin{array}{ccc}j_{12}&J_B&J_E\\-z_e&j_e&-m_e\end{array}\right)
(-1)^{z_e-z_a}D^{j_{12}}_{z_az_e}(U_{12}).
\la{decomp}\eeq

We now combine together $3jm$ symbols related to the same vertices
(they are marked by appropriate indices of the projections
$a,b,c,d,r,f,g,h$), three $3jm$ symbols for each vertex, together with
appropriate sign factors. The three $3jm$ symbols per vertex combine
into $6j$ symbols, one for each vertex of the cube.

\vskip .3true cm

\underline{Vertex $a$}

\vskip .3true cm

Related to vertex $a$ are the factors

\[
\sum_{i_a,j_a,m_a}(-1)^{i_a+j_a+m_a-o_a-r_a-z_a}
\left(\begin{array}{ccc}j_1&J_A&J_B\\-o_a&i_a&-j_a\end{array}\right)
\left(\begin{array}{ccc}j_4&J_E&J_A\\-r_a&m_a&-i_a\end{array}\right)
\left(\begin{array}{ccc}j_{12}&J_B&J_E\\-z_a&j_a&-m_a\end{array}\right).
\nonumber\]
[we use $o_a+r_a+z_a=0$, make cyclic permutations
in all $3jm$ symbols, and change the summation indices $i,j,m\to
-i,-j,-m$]
\[
=\sum_{i_a,j_a,m_a}(-1)^{-i_a-j_a-m_a}
\left(\begin{array}{ccc}J_B&j_1&J_A\\j_a&-o_a&-i_a\end{array}\right)
\left(\begin{array}{ccc}J_A&j_4&J_E\\i_a&-r_a&-m_a\end{array}\right)
\left(\begin{array}{ccc}J_E&j_{12}&J_B\\m_a&-z_a&-j_a\end{array}\right)
\]
\beq
=(-1)^{-J_A-J_B-J_E}
\left(\begin{array}{ccc}j_1&j_4&j_{12}\\o_a&r_a&z_a\end{array}\right)
\left\{\begin{array}{ccc}j_1&j_4&j_{12}\\J_E&J_B&J_A\end{array}\right\}.
\la{a}\eeq
In the last transformation the definition of the $6j$ symbol
\ur{6jdef1} has been used.

\vskip .3true cm

\underline{Vertex $b$}

\vskip .3true cm

Related to vertex $b$ are the factors

\[
\sum_{i_b,j_b,k_b}\left.(-1)^{o_b+k_b}\right|_{o_b=i_b-j_b}
\left(\begin{array}{ccc}j_1&J_A&J_B\\-o_b&i_b&-j_b\end{array}\right)
\left(\begin{array}{ccc}j_2&J_A&J_C\\-p_b&i_b&-k_b\end{array}\right)
\left(\begin{array}{ccc}j_9&J_C&J_B\\-w_b&k_b&-j_b\end{array}\right).
\nonumber\]
[we interchange the first two columns in the first $3jm$ symbol
and change the signs of all its projections;
it doesn't change the sign of the $3jm$'s. Also, we
make cyclic permutations of the last two $3jm$ symbols,
and change the summation indices $i,j,k\to -i,-j,-k$]

\[
=\sum_{i,j,k}(-1)^{-i+j-k}\left[{\rm insert}\;1=(-1)^{2J_B-2j}\right]
\]
\[\cdot
\left(\begin{array}{ccc}J_A&j_1&J_B\\i&o_b&-j\end{array}\right)
\left(\begin{array}{ccc}J_B&j_4&J_C\\j&-w_b&-k\end{array}\right)
\left(\begin{array}{ccc}J_C&j_9&J_A\\k&-p_b&-i\end{array}\right)
\]
\beq
=(-1)^{J_B-J_A-J_C}
\left(\begin{array}{ccc}j_1&j_9&j_2\\-o_b&w_b&p_b\end{array}\right)
\left\{\begin{array}{ccc}j_1&j_9&j_2\\J_C&J_A&J_B\end{array}\right\}.
\la{b}\eeq

In each case we combine the three $3jm$ symbols and the sign
factors so that they suit the definition of the $6j$ symbol given in
Appendix A, \eq{6jdef1}.

An important property of the sign factors is the following:
if $j_1,J_A,J_B$ enter one $3jm$ symbol, there is an equality:

\beq
(-1)^{\pm 2j_1\pm 2J_A\pm 2J_B}=1,
\la{signf}\eeq
where all signs are possible. This is because out of three momenta
either zero or two moments are half-integer. Another important property
is that, if $J$ is the momentum entering a certain $3jm$ symbol, and
$m$ is its projection, then $(-1)^{2J\pm 2m}=+1$. This is because
$J$ and $m$ are either integer or half-integer, but simultaneously.

Below we cite without detailed derivation (which is quite similar
to those above) the expressions for other vertices of the cube.

\vskip .3true cm

\underline{Vertex $c$}

\vskip .3true cm

\beq
=(-1)^{J_A+J_D-J_C}
\left(\begin{array}{ccc}j_2&j_3&j_{10}\\p_c&q_c&-x_c\end{array}\right)
\left\{\begin{array}{ccc}j_2&j_3&j_{10}\\J_D&J_C&J_A\end{array}\right\}.
\la{c}\eeq

\vskip .3true cm

\underline{Vertex $d$}

\vskip .3true cm

\beq
=(-1)^{J_A-J_D-J_E}
\left(\begin{array}{ccc}j_4&j_3&j_{11}\\-r_d&q_d&y_d\end{array}\right)
\left\{\begin{array}{ccc}j_4&j_3&j_{11}\\J_D&J_E&J_A\end{array}\right\}.
\la{d}\eeq

\vskip .3true cm

\underline{Vertex $e$}

\vskip .3true cm

\beq
=(-1)^{J_E-J_B-J_F}
\left(\begin{array}{ccc}j_{12}&j_8&j_5\\-z_e&v_e&s_e\end{array}\right)
\left\{\begin{array}{ccc}j_{12}&j_8&j_5\\J_F&J_B&J_E\end{array}\right\}.
\la{e}\eeq

\vskip .3true cm

\underline{Vertex $f$}

\vskip .3true cm

\beq
=(-1)^{J_B+J_C-J_F}
\left(\begin{array}{ccc}j_6&j_5&j_9\\-t_f&s_f&w_f\end{array}\right)
\left\{\begin{array}{ccc}j_6&j_5&j_9\\J_B&J_C&J_F\end{array}\right\}.
\la{f}\eeq

\vskip .3true cm

\underline{Vertex $g$}

\vskip .3true cm

\beq
=(-1)^{J_C+J_D+J_F}
\left(\begin{array}{ccc}j_6&j_{10}&j_7\\t_g&x_g&u_g\end{array}\right)
\left\{\begin{array}{ccc}j_6&j_{10}&j_7\\J_D&J_F&J_C\end{array}\right\}.
\la{g}\eeq

\vskip .3true cm

\underline{Vertex $h$}

\vskip .3true cm

\beq
=(-1)^{J_E+J_F-J_D}
\left(\begin{array}{ccc}j_7&j_{11}&j_8\\-u_h&y_h&v_h\end{array}\right)
\left\{\begin{array}{ccc}j_7&j_{11}&j_8\\J_E&J_F&J_D\end{array}\right\}.
\la{h}\eeq

Combining all these factors we get \eq{cube2} corresponding to the
cube.

\section*{Appendix C. 6j symbols at the lattice sites}

In this Appendix we show how integration over link variables in
\eq{cube2} combine, together with the $3jm$ factors, into $6j$
symbols composed of the link momenta $j$, one for each site of the
lattice. The notations are given in Fig.3.

Let us consider integration over link variables $U_{1,4,12,13,14,15}$
entering the vertex $a$ shown in Fig.3. This vertex is an intersection
of four even cubes denoted in Fig.3 as $I$, $II$, $III$ and $IV$.
Link $1$ is common to the cubes $I$ and $II$, link $4$ is common to $I$
and $IV$, and so on.

The analytical expression for the cube $I$ is given by \eq{cube2}.
The factors relevant to vertex $a$ are

\beq
D^{j_1}_{o_ao_b}(U_1)D^{j_4}_{r_ar_d}(U_4) D^{j_{12}}_{z_az_e}(U_{12})
\left(\begin{array}{ccc}j_1&j_4&j_{12}\\o_a&r_a&z_a\end{array}\right).
\la{Ia}\eeq

It is not necessary to compute anew corresponding expressions for the
cubes $II-IV$. It is sufficient to draw a correspondence between the
links and the sites of other cubes with those of the cube $I$. For
example, link $1$, as seen from the viewpoint of cube $II$, is analogous
to link $7$ of cube $I$; the vertex $a$ from the viewpoint of cube $II$
is analogous to vertex $h$ of cube $I$, and vertex $b$ is analogous to
vertex $g$. In the table below we give the list of the `analogs' of
links in cubes $II-IV$ to those of the cube $I$.

\vskip .3true cm

\begin{center}
\begin{tabular}{|c|c|}
\hline
II & I \\
\hline
\hline
1&7\\
\hline
13& 11 \\
\hline
14 & 8\\
\hline
a & h \\
\hline
\end{tabular}
\hskip 1true cm
\begin{tabular}{|c|c|}
\hline
III & I \\
\hline
\hline
12&10\\
\hline
14& 2 \\
\hline
15 & 3\\
\hline
a & c \\
\hline
\end{tabular}
\hskip 1true cm
\begin{tabular}{|c|c|}
\hline
IV & I \\
\hline
\hline
4&6\\
\hline
13& 9 \\
\hline
15 & 5\\
\hline
a & f \\
\hline
\end{tabular}
\end{center}

Having this table of correspondence we can immediately read off from
\eq{cube2} the expressions relevant to the vertex $a$, arising from the
cubes $II-IV$:

\beq
{\rm from\;cube\;}II:\qquad
D^{j_1^\prime}_{u_au_b}(U_1)D^{j_{13}}_{-y_a,-y_\beta}(U_{13}^\dagger)
D^{j_{14}}_{-v_a,-v_\alpha}(U_{14}^\dagger)
\left(\begin{array}{ccc}j_1^\prime&j_{13}&j_{14}\\-u_a&y_a&v_a
\end{array}\right),
\la{IIa}\eeq

\beq
{\rm from\;cube\;}III:\qquad
D^{j_{12}^\prime}_{x_ax_e}(U_{12})D^{j_{14}^\prime}_{-p_a,-p_\alpha}
(U_{14}^\dagger) D^{j_{15}}_{-q_a,-q_\epsilon}(U_{15}^\dagger)
\left(\begin{array}{ccc}j_{12}^\prime&j_{14}^\prime&j_{15}\\-x_a&p_a&q_a
\end{array}\right),
\la{IIIa}\eeq

\beq
{\rm from\;cube\;}IV:\qquad
D^{j_4^\prime}_{u_au_b}(U_4)D^{j_{15}^\prime}_{-s_a,-s_\epsilon}
(U_{15}^\dagger) D^{j_{13}^\prime}_{-w_a,-w_\beta}(U_{13}^\dagger)
\left(\begin{array}{ccc}j_4^\prime&j_{15}^\prime&j_{13}^\prime\\
-t_a&s_a&w_a\end{array}\right).
\la{IVa}\eeq

Integrating over $U_{1,4,12,13,14,15}$ we get:

\bea
\int dU_1 D^{j_1}_{o_ao_b}(U_1)D^{j_1^\prime}_{u_au_b}(U_1)
&=&\frac{\delta_{j_1j_1^\prime}}{2j_1+1}(-1)^{u_b-u_a}
\delta_{o_a,-u_a}\delta_{o_b,-u_b},
\la{U1}\\
\int dU_4 D^{j_4}_{r_ar_d}(U_4)D^{j_4^\prime}_{t_at_d}(U_4)
&=&\frac{\delta_{j_4j_4^\prime}}{2j_4+1}(-1)^{t_d-t_a}
\delta_{r_a,-t_a}\delta_{r_d,-t_d},
\la{U4}\\
\int dU_{12}
D^{j_{12}}_{z_az_e}(U_{12})D^{j_{12}^\prime}_{x_ax_e}(U_{12})
&=&\frac{\delta_{j_{12}j_{12}^\prime}}{2j_{12}+1}(-1)^{x_e-x_a}
\delta_{z_a,-x_a}\delta_{z_e,-x_e},
\la{U12}\\
\int dU_{13}
D^{j_{13}}_{-y_a,-y_\beta}(U_{13}^\dagger)
D^{j_{13}^\prime}_{-w_a,-w_\beta}(U_{13}^\dagger)
&=&\frac{\delta_{j_{13}j_{13}^\prime}}{2j_{13}+1}(-1)^{w_a-w_\beta}
\delta_{w_a,-y_a}\delta_{w_\beta,-y_\beta},
\la{U13}\\
\int dU_{14}
D^{j_{14}}_{-v_a,-v_\alpha}(U_{14}^\dagger)
D^{j_{14}^\prime}_{-p_a,-p_\alpha}(U_{14}^\dagger)
&=&\frac{\delta_{j_{14}j_{14}^\prime}}{2j_{14}+1}(-1)^{p_a-p_\alpha}
\delta_{p_a,-v_a}\delta_{p_\alpha,-v_\alpha},
\la{U14}\\
\int dU_{15}
D^{j_{15}}_{-q_a,-q_\epsilon}(U_{15}^\dagger)
D^{j_{15}^\prime}_{-s_a,-s_\epsilon}(U_{15}^\dagger)
&=&\frac{\delta_{j_{15}j_{15}^\prime}}{2j_{15}+1}(-1)^{s_a-s_\epsilon}
\delta_{s_a,-q_a}\delta_{s_\epsilon,-q_\epsilon}.
\la{U15}\eea

The four $3jm$ symbols in eqs.(\ref{Ia}-\ref{IVa}) get now fully
contracted over all indices. This results in a $6j$ symbol according
to \eq{6jdef2} of Appendix A. Indeed we have for vertex $a$:

\[
{\rm ``a"}=\sum_{orqvyz}(-1)^{o+r+z-q-v-y}\]
\beq
\left(\begin{array}{ccc}j_1&j_4&j_{12}\\o&r&z\end{array}\right)
\left(\begin{array}{ccc}j_1&j_{13}&j_{14}\\o&y&v\end{array}\right)
\left(\begin{array}{ccc}j_{12}&j_{14}&j_{15}\\z&-v&q\end{array}\right)
\left(\begin{array}{ccc}j_4&j_{15}&j_{13}\\r&-q&-y\end{array}\right)
\nonumber\eeq
[we note that $o+r+z=0$; we change the summation variable $y\to -y$,
and interchange the last two columns in the second $3jm$ symbol and
the first two columns in the last two $3jm$ symbols -- that gives
sign factors $(-1)^{j_1+j_{13}+j_{14}}$, $(-1)^{j_{12}+j_{14}+j_{15}}$
and $(-1)^{j_4+j_{13}+j_{15}}$. Finally, we insert two unities in the
form of $1=(-1)^{2v-2j_{14}}$ and $1=(-1)^{2q-2j_{15}}$]

\[
=(-1)^{j_1+j_4+j_{12}+j_{13}-j_{14}-j_{15}}\sum_{orqvyz}
\]
\[
\left(\begin{array}{ccc}j_1&j_4&j_{12}\\o&r&z\end{array}\right)
\left(\begin{array}{ccc}j_1&j_{14}&j_{13}\\o&v&-y\end{array}\right)
\left(\begin{array}{ccc}j_{15}&j_{4}&j_{13}\\-q&r&y\end{array}\right)
\left(\begin{array}{ccc}j_{15}&j_{14}&j_{12}\\q&-v&z\end{array}\right)
\]
\beq
=(-1)^{j_1+j_4+j_{12}+j_{13}-j_{14}-j_{15}}
\left\{\begin{array}{ccc}j_1&j_4&j_{12}\\j_{15}&j_{14}&j_{13}
\end{array}\right\}
\nonumber\eeq
[since $j_{12}$, $j_{14}$ and $j_{15}$ came from one $3jm$ symbol
one can use the equation (see \eq{signf}) $(-1)^{j_{12}-j_{14}-j_{15}}=
(-1)^{-j_{12}+j_{14}+j_{15}}$]
\beq
=(-1)^{j_1+j_4-j_{12}+j_{13}+j_{14}+j_{15}}
\left\{\begin{array}{ccc}j_1&j_4&j_{12}\\j_{15}&j_{14}&j_{13}
\end{array}\right\}.
\la{verta}\eeq

This is the final result for the vertex $a$: the six angular momenta
ascribed to the six links entering this vertex combine to produce a
$6j$ symbol.

Similarly, one can treat the vertex $b$, see Fig.3. Links labelled by
numbers $1,2,9,16,17,18$ enter this vertex; they are pair-wise shared
by the cubes $I,II,V$ and $VI$. The correspondence between the links
viewed from the viewpoint of the cubes $II,V,VI$ with those of the cube
$I$ is given by the following table:

\vskip .3true cm

\begin{center}
\begin{tabular}{|c|c|}
\hline
II & I \\
\hline
\hline
1&7\\
\hline
16& 6 \\
\hline
18 & 10\\
\hline
b & g \\
\hline
\end{tabular}
\hskip 1true cm
\begin{tabular}{|c|c|}
\hline
V & I \\
\hline
\hline
9 &11\\
\hline
16& 4 \\
\hline
17 & 3\\
\hline
b & d \\
\hline
\end{tabular}
\hskip 1true cm
\begin{tabular}{|c|c|}
\hline
VI & I \\
\hline
\hline
2&8\\
\hline
17& 5 \\
\hline
18 & 12\\
\hline
b & e \\
\hline
\end{tabular}
\end{center}

Performing the same steps as in deriving the $6j$ symbol for the vertex
$a$ we arrive to the following result for the vertex $b$:

\beq
{\rm ``b"}\;=(-1)^{j_1+j_2+j_9+j_{16}+j_{17}-j_{18}}
\left\{\begin{array}{ccc}j_1&j_9&j_2\\j_{17}&j_{18}&j_{16}
\end{array}\right\}.
\la{vertb}\eeq

We notice that vertex $a$ is of the `even' and vertex $b$ is of the
`odd' type: all other vertices of the lattice can be considered as
either  `even' or `odd'. Therefore, \eqs{verta}{vertb} give actually
the full result. Combining them together we find that a sign factor

\beq
(-1)^{2j}=(-1)^{-2j}
\la{sf1}\eeq
should be attributed to {\em all} links of the lattice.

Let us prove that this sign factor is equivalent (in the vacuum!)
to a sign factor

\beq
(-1)^{2J}=(-1)^{-2J}
\la{sf2}\eeq
attributed to {\em all} plaquettes of the lattice. We recall
that all links are shared by two even cubes whose faces
carry plaquette values $J$. We first attribute all links to only one
(out of the two possible) cubes, according to some rule. Many such
rules can be suggested, the only requirement being that each link
is attributed to one and only one even cube. An example is given by
the following construction: we choose the edges 12,5,9,2 and 7
(see Fig.2) as `belonging' to the cube shown on that figure. The rest
six edges will then `belong' to one of the neighbouring even cubes.
For example, the edge 1 will be counted as `belonging' to the cube II
(see Fig.3). Indeed, from the cube II point of view that edge will be
of the type 7, and so forth. It can be seen that, in these scheme,
every link of the full lattice will `belong' to one and only one
even cube.

We have, therefore, a sign factor
\beq
(-1)^{2j_{12}+2j_5+2j_9+2j_2+2j_7}
\la{signf3}\eeq
attributed to the cube I. Next, we recall that, e.g., $j_{12}$ enters
the $3jm$ symbol together with the plaquette angular momenta $J_B$ and
$J_E$ (see \ur{a}). Using \eq{signf} appropriate to the case we can
replace $(-1)^{2j_{12}}=(-1)^{2J_B+2J_E}$. Similarly,
$(-1)^{2j_5}=(-1)^{2J_B+2J_F}$, and so on. As a result we get that the
sign factor \ur{signf3} is equal to

\beq
(-1)^{2J_A+2J_B+2J_C+2J_D+2J_E+2J_F}.
\la{signf4}\eeq
This procedure can be repeated for all even cubes of the lattice.
It proves the above statement that the product of all link sign factors
\ur{sf1} can be replaced by the product of all plaquette sign factors
\ur{sf2}. It should be stressed that this proof is valid only for the
vacuum, i.e. for the partition function itself but, generally speaking,
not for the averages of operators.

\section*{Appendix D. 9j symbols from the Wilson loop}

Let the Wilson loop in the representation $j_s$ go through the links
...,15,1,17,..., see Fig.3 for notations. It means that one has now to
integrate three $D$-functions of the link variables $U_{15,1,...}$,
instead of two, as it was in \eqs{U1}{U15} of the previous Appendix,
the rest integrations remaining unchanged. We have now

\beq
\int dU_1 D^{j_1}_{o_ao_b}(U_1)D^{j_1^\prime}_{u_au_b}(U_1)
D^{j_s}_{m_am_b}(U_1)
=\left(\begin{array}{ccc}j_1&j_1^\prime&j_s\\o_a&u_a&m_a\end{array}\right)
\left(\begin{array}{ccc}j_1&j_1^\prime&j_s\\o_b&u_b&m_b\end{array}\right),
\la{U1W}\eeq
\[
\int dU_{15}
D^{j_{15}}_{-q_a,-q_\epsilon}(U_{15}^\dagger)
D^{j_{15}^\prime}_{-s_a,-s_\epsilon}(U_{15}^\dagger)
D^{j_s}_{m_bm_\epsilon}(U_{15})
\]
\beq
=(-1)^{m_\epsilon-m_a}
\left(\begin{array}{ccc}j_{15}&j_{15}^\prime&j_s\\
q_\epsilon&s_\epsilon&m_\epsilon\end{array}\right)
\left(\begin{array}{ccc}j_{15}&j_{15}^\prime&j_s\\
q_a&s_a&m_a\end{array}\right).
\la{U15W}\eeq

Using the other $3jm$ symbols related to the vertex $a$ (see
eqs.(\ref{Ia}-\ref{IVa})) and the Kronecker symbols from
eqs.(\ref{U4}-\ref{U14}) we get for the vertex $a$:

\[
{\rm ``a"}=\sum(-1)^{r+z+w+p-m}
\left(\begin{array}{ccc}j_1&j_4&j_{12}\\o&r&z\end{array}\right)
\left(\begin{array}{ccc}j_1^\prime&j_{13}&j_{14}\\-u&-w&-p\end{array}\right)
\]
\beq
\left(\begin{array}{ccc}j_{14}&j_{15}&j_{12}\\p&q&z\end{array}\right)
\left(\begin{array}{ccc}j_4&j_{15}^\prime&j_{13}\\r&s&w\end{array}\right)
\left(\begin{array}{ccc}j_1&j_1^\prime&j_s\\o&u&m\end{array}\right)
\left(\begin{array}{ccc}j_{15}&j_{15}^\prime&j_s\\q&s&m\end{array}\right)
\nonumber\eeq
[we notice that $r+z=-o,\;w+p=-u$ and that $o+u+m=0$, hence the sign
factor is $+1$; we change the signs of all projections in the second
$3jm$ symbol, and permute the columns in other $3jm$ symbols to match
the definition of the $9j$ symbols as given by \eq{9jdef}]

\beq
=(-1)^{j_1^\prime-j_4+j_{14}+j_{15}+j_s}
\left\{\begin{array}{ccc}j_4&j_1&j_{12}\\j_{15}^\prime&j_s&j_{15}\\
j_{13}&j_1^\prime&j_{14}\end{array}\right\}.
\la{9ja}\eeq
To get the final sign factor we have used the relation $(-1)^{\pm
2j_1\pm 2j_2\pm 2j_3}=+1$ valid for any $j_{1,2,3}$ originating from
one $3jm$ symbol.

Acting in the same fashion we obtain for the vertex $b$:

\[
{\rm ``b"}=\sum (-1)^{-v-y-t-x+m}
\left(\begin{array}{ccc}j_1&j_9&j_2\\-o&-y&-v\end{array}\right)
\left(\begin{array}{ccc}j_{16}&j_{18}&j_1^\prime\\t&x&u\end{array}\right)
\]
\[
\left(\begin{array}{ccc}j_{16}&j_{17}&j_9\\t&q&y\end{array}\right)
\left(\begin{array}{ccc}j_{18}&j_2&j_{17}^\prime\\x&v&s\end{array}\right)
\left(\begin{array}{ccc}j_1&j_1^\prime&\j_s\\o&u&m\end{array}\right)
\left(\begin{array}{ccc}j_{17}&j_{17}^\prime&j_s\\q&s&m\end{array}\right)
\]
\beq
=(-1)^{j_1^\prime-j_2+j_{16}+j_{17}+j_s}
\left\{\begin{array}{ccc}j_2&j_1&j_9\\j_{17}^\prime&j_s&j_{17}\\
j_{18}&j_1^\prime&j_{16}\end{array}\right\}.
\la{9jb}\eeq

\end{document}